\documentclass[a4paper,fleqn]{cas-sc}

\usepackage[numbers]{natbib}
\usepackage{lineno}

\usepackage{multirow}
\usepackage{adjustbox}

\def\tsc#1{\csdef{#1}{\textsc{\lowercase{#1}}\xspace}}
\tsc{WGM}
\tsc{QE}

\let\oldequation\equation
\let\oldendequation\endequation

\renewenvironment{equation}
  {\linenomathNonumbers\oldequation}
  {\oldendequation\endlinenomath}

\begin{document}
\let\WriteBookmarks\relax
\def\floatpagepagefraction{1}
\def\textpagefraction{.001}

\shorttitle{}    

\shortauthors{}  

\title [mode = title]{Ultrasonic backscattering model for Rayleigh waves in polycrystals with Born and independent scattering approximations}  



%

\author[1,2]{Shan Li}



\ead{lisa_13@foxmail.com}

\affiliation[1]{organization={School of Traffic and Transportation Engineering},
            addressline={Central South University}, 
            city={Chang Sha},
            postcode={410075}, 
            state={Hunan},
            country={China}}
            
\affiliation[2]{organization={Department of Mechanical Engineering},
            addressline={Imperial College London}, 
            city={Exhibition Road},
            postcode={SW7 2AZ}, 
            state={London},
            country={United Kingdom}}

\author[2]{Ming Huang}
\ead{m.huang16@imperial.ac.uk}

\author[1]{Yongfeng Song}
\cormark[1]
\ead{songyf_ut@csu.edu.cn}

\author[2]{Bo Lan}
\cormark[1]
\ead{bo.lan@imperial.ac.uk}

\author[1]{Xiongbing Li}
\ead{lixb213@csu.edu.cn}
\begin{abstract}
This paper presents theoretical and numerical models for the backscattering of 2D Rayleigh waves in single-phase, untextured polycrystalline materials with statistically equiaxed grains. The theoretical model, based on our prior inclusion-induced Rayleigh wave scattering model and the independent scattering approximation, considers single scattering of Rayleigh-to-Rayleigh (R-R) waves. The numerical finite element model is established to accurately simulate the scattering problem and evaluate the theoretical model. Good quantitative agreement is observed between the theoretical model and the finite element results, especially for weakly scattering materials. The agreement decreases with the increase of the anisotropy index, owing to the reduced applicability of the Born approximation. However, the agreement remains generally good when weak multiple scattering is involved. In addition, the R-R backscattering behaviour of 2D Rayleigh waves is similar to the longitudinal-to-longitudinal and transverse-to-transverse backscattering of bulk waves, with the former exhibiting stronger scattering. These findings establish a foundation for using Rayleigh waves in quantitative characterisation of polycrystalline materials.
\end{abstract}




\maketitle


\section{Introduction}\label{Sec.1}
Rayleigh waves, when propagating on the surface of a polycrystalline material, can be scattered by the grain boundaries due to the acoustic impedance contrast caused by different alignments of crystallographic orientations of individual grains \citep{ryzy2018measurement,grabec2022surface}. The backscattered waves - the portion of the scattered wave that travels back to the transducer - are sometimes called backscattered `grain noise', and this phenomenon of the bulk wave has been the subject of thorough scientific investigations. For example, backscattered waves are widely proven to be able to characterise the material's microstructure and estimate material properties, e.g. the size of the grains \citep{margetan1994backscattered}, the degree of preferred texture \citep{han1996effect}, and the multiphase content \citep{rose1993theory}. 

In the past decades, there have been several physical quantities, including the figure of merit (FOM)  \citep{margetan1994backscattered,margetan1993detectability}, also called the backscattering coefficient   \citep{rose1993theory,rose1991ultrasonic,rose1992ultrasonic}, and theoretical models developed in order to quantify the backscattered amplitude and intensity of scattered energy. The theoretical models include  the independent scattering model (ISM)  \citep{margetan1991technique, margetan1993modeling},  singly scattered response (SSR) \citep{ghoshal2007wigner,ghoshal2010diffuse, hu2013mode,hu2017transverse} and doubly scattered response (DSR) \citep{hu2015contribution,huang2021transverse} for different bulk wave types. Meanwhile, based on these models, research has been performed to study materials with more complicated microstructures \citep{lobkis2012ultrasonic,yang2013ultrasonic,arguelles2016mode}. Here, we recognise that compared with other models mentioned, the model describing FOM is advantageous for the characterisation of the microstructure of material from a practical point of view, because it involves a simple mathematical expression and does not need the consideration of experimental conditions.

While comprehensive models have been established for the scattering of bulk waves by grains, limited studies have focused on the scattering of Rayleigh waves in polycrystalline materials. Zhang and Weaver studied the singly incoherent scattered field of leaky Rayleigh waves from a fluid/solid surface at the  critical Rayleigh angle using the first Born approximation \citep{zhang1996leaky}. They proposed that the mean-square scattered signal level is given in terms of integration of the spatial-spectral density (the spatial Fourier transform of the autocovariance function of the fluctuating elastic moduli). Except for the research of the mean-square scattered signal level, scattering attenuation of Rayleigh waves has also been investigated. For example, Kaganova and Maradudin gave an expression of the scattering attenuation and dispersion relationship of Rayleigh waves  \citep{kaganova1992surface}. However, their theoretical model is difficult to solve, and no quantitative results have been obtained. Recently, Ryzy \textit{et al.} \citep{ryzy2018measurement} and Li \textit{et al.} \citep{li2022attenuation} predicted the scattering attenuation for different types of Rayleigh waves. However, these cases are only interested in the scattering attenuation. The explicit expression for the backscattered signal of Rayleigh waves is not given yet now. 

Given Rayleigh waves' capability to quantitatively evaluate the material properties in near-surface regions, we consider it of considerable interest to develop the theory needed to describe Rayleigh wave backscattering behaviour from a polycrystalline material. Lately, we have given an explicit expression for the backward flaw scattering amplitude of R-R scattered by a single inclusion (weak scatterer) based on Born approximation \citep{li2022theoretical}, which would provide theoretical support for the backscattered grain noise research. With the independent scattering (IS) approximation \citep{rose1991ultrasonic,margetan1991technique,margetan1993modeling}, the total backscattering power of Rayleigh waves can be interpreted as an incoherent sum of the power scattered from each grain. Thus, an opportunity exists to employ a theoretical method, based on the Born and IS approximations, to complement our understanding of the backscattering grain noise of Rayleigh waves when propagating on the plain surface of polycrystalline materials with single-phase, untextured and equiaxed grains.

In addition to the theoretical methods, finite element (FE) modelling is another approach which has been widely used to investigate wave scattering behaviours in a polycrystalline material. Similar to the theoretical developments in literature, the related works in this area have also mainly concentrated on bulk wave grain noise, with two-dimensional (2D)  \citep{bai2018finite,ghoshal2009diffuse} and three-dimensional (3D) models \citep{liu2019investigation} both well researched. Recently, there have also been successful applications of FE in analysing the scattering attenuation and velocity dispersion of Rayleigh waves in the polycrystalline material  \citep{grabec2022surface,ryzy2020finite}. Compared to experiments where significant limitations on the testing conditions and knowledge of the detailed materials microstructures are present, these numerical studies allow full control and knowledge of the materials, demonstrating the power of the FE method as a perfectly controlled experiment to realistically simulate Rayleigh wave scattering. Therefore, in this paper, we combine the development of new theoretical advancements with powerful FE simulations for verification purposes.

In comparison to the existing studies, this work sets out to study 2D Rayleigh wave grain backscattering behaviour. To achieve this aim, this work contributes to two aspects.  (1) The work develops explicit formulae for the backscattered power from single-phase, untextured, and equiaxed grains based on the Born and IS approximations, which leads to the calculation of the backscattered grain noise of Rayleigh waves. (2) We make use of the proven capability of the FE method to perform realistic simulations of backward scattering amplitudes of Rayleigh waves scattered by grains of a polycrystal. This not only allows Rayleigh wave scattering behaviour to be studied numerically as a standalone application, some of the outputs can also be input back into the theoretical model to perform numerical integration, thus allowing thorough verification of the theoretical model. 

The paper is organised to explain the methodology and highlight the contributions clearly: Section \ref{sec:2} presents the theoretical model to explain the backward scattering behaviours of the Rayleigh wave based on the approximations. Then Sec. \ref{sec:3} gives a brief introduction to an FE model which illustrates the R-R wave responses after being scattered by grains. The comparisons between the theoretical and computational results are shown in Sec. \ref{sec:4}, mainly including that the measured single scattering amplitudes are scattered from single grain with different shapes and random orientation, shown in Sec. \ref{sec:4.1}; the theoretical model is verified to evaluate the root-mean-square (rms) backward grain noise with different anisotropy index materials in Sec. \ref{sec:4.2}. Finally, conclusions are given in Sec. \ref{sec:5}.

\section{\label{sec:2} Backscattering of Rayleigh waves}

\subsection{\label{sec:2.1} Brief review for the inclusion scattering of Rayleigh waves}
The interest here is studying the Rayleigh wave propagation on the smooth polycrystal plain surface. Usually, the grain can be regarded as a weak scatter with a small elastic constant perturbation. Therefore, the scattering behaviour from the single grain can be simplified to the inclusion scattering. In fact, the theoretical model related to the inclusion scattering of Rayleigh waves has been well established by our previous research  \citep{li2022theoretical}, which is developed based on the Born approximation in which the displacement fields for the scattered wave are approximated by those of the incident wave.

Now, a brief overview of the inclusion backscattering theory of the R-R waves is introduced here. We consider a statistically isotropic solid as the host material with constant density $\rho$ in the two-dimensional (2D) half-space defined by the $x-z$ coordinates. An arbitrarily-shaped inclusion is present on the surface or subsurface of the host material. The inclusion is defined in the region $V$.  The anisotropic property of the host material and inclusion are described by the elastic tensor $C_{pjkl}^0$ and $C_{pjkl}^1 \left(\mathbf{x}_{\mathrm{s}}\right)$. The anisotropic property difference between the inclusion and host material is described by the elastic tensor $\Delta C_{pjkl} \left(\mathbf{x}_{\mathrm{s}}\right) = C_{pjkl}^1 \left(\mathbf{x}_{\mathrm{s}}\right)-C_{pjkl}^0 $. In addition, there is no density change caused by the inclusion.
Based on the reciprocity theorem and the Born approximation, the backscattered Rayleigh wave can be given as \citep{li2022theoretical},
\begin{equation}
\begin{split}
   {u}^\text{sc}_{n} \left(\mathbf{x}, \omega \right)  \approx \int_{V} &\left [ 
    -\Delta C_{pjkl}(\mathbf{x}_{\mathrm{s}})  G_{ni, j} \left(\mathbf{x},\mathbf{x}_{\mathrm{s}}, \omega \right) {u}_{k, l}^\text{in} \left(\mathbf{x}_{\mathrm{s}}, \omega \right) \right ]  \mathrm{~d} V \textrm{.}
\label{eq:001}
\end{split}
\end{equation}
The ${u}_{k, l}^\text{in} \left(\mathbf{x}_{\mathrm{s}}, \omega \right) $ is the derivative of the incident Rayleigh wave displacement, given as,
\begin{equation}
        {u}_{k, l}^\text{in}(\mathbf{x}_{\mathrm{s}}) = \left[{d}_{k, l}^\text{in}(z_\mathrm{s}) + \text{i} {d}_{k}^\text{in}(z_\mathrm{s}) k_R {e}^\text{in}_{l} \right] \exp{\left(\text{i} k_R \mathbf{e}^\text{in} \cdot \mathbf{x}_{\mathrm{s}} \right)}\textrm{,} 
        \label{eq:displacement}
\end{equation}
with
\begin{equation}
\begin{split}
  &\quad \quad \quad \quad \quad \quad \quad \quad \quad \quad \quad \quad {d}^\text{in}_{k}  = \left[U_{R}(z_\mathrm{s}) , 0 , W_{R}(z_\mathrm{s})\right]\textrm{,}  \\
 &U_{R}\left(z\right) =  w_1^{L}
 \exp \left(-\eta_{L} z\right)+w_1^{T}
 \exp \left(-\eta_{T} z\right),  \quad W_{R}\left(z\right)  =  w_3^{L}
 \exp \left(-\eta_{L} z\right)  +w_3^{T}
 \exp \left(-\eta_{T} z\right), \\
&w_{1}^{L}   =  k_R\left(2 c_{{T}}^2-c_R^2\right)\big /2 \eta_L c_T^2,  \quad w_{1}^{T}   =  - {\eta_T}\big /{k_R}, \quad w_{3}^{L} = \text{i}{\left(2 c_T^2-c_R^2\right)}\big /{2 c_T^2 }, \quad w_{3}^{T} =  - 1\text{i},  \\
&\quad \quad \quad \quad \quad \quad \quad \quad \eta_{L} = k_R \sqrt{1-c_R^2\big /c_L^2},  \quad
        \eta_{T} = k_R \sqrt{1-c_R^2\big /c_T^2},
    \end{split}
\end{equation}
where  $k_R$ and $c_R$ are the wave number and phase velocity of the incident Rayleigh wave. $\mathbf{e}^\text{in} = [1,0,0]$ is the propagation direction of the incident Rayleigh wave. The phase velocity $c_R$ can be calculated by \citep{viktrov1967rayleigh},
\begin{equation}
\label{eq:04}
 \left(2-c_R^2\big /c_T^2\right)^2-4\left(1-c_R^2\big /c_L^2\right)^{1/2}\left(1-c_R^2\big /c_T^2\right)^{1/2} = 0 \textrm{,}
\end{equation}
where $c_L$ and $c_T$ are the Voigt-averaged velocities of the longitudinal and shear waves in the host material, which can be obtained by \citep{kube2015voigt},
\begin{equation}
    c_L = \sqrt{ c_{11}^0  \big / \rho}, \quad c_T = \sqrt{ c_{44}^0  \big / \rho},
\end{equation}
where $c_{11}^{0}$ and $c_{44}^{0}$ are the Voigt-averaged constants, which can be  given by

\begin{equation}\label{eq:voigt}
\begin{split}
 c_{11}^0 &= \frac{3\left(c_{11}+c_{22}+c_{33}\right)+2\left(c_{23}+c_{13}+c_{12}\right)+4\left(c_{44}+c_{55}+c_{66}\right)}{15} \\
 c_{44}^0 &= \frac{\left(c_{11}+c_{22}+c_{33}\right)-\left(c_{23}+c_{13}+c_{12}\right)+3\left(c_{44}+c_{55}+c_{66}\right)}{15} \textrm{,}
\end{split}
\end{equation}
The fourth-rank elastic tensor $C_{pjkl}$ is written as $c_{ij}$ using the Voigt index notation where the pairs of indices are contracted to the following single values: $11\rightarrow 1$, $22\rightarrow 2$, $33\rightarrow 3$, $23\,\mathrm{or}\,32\rightarrow 4$, $13\,\mathrm{or}\,31\rightarrow 5$ and $12\,\mathrm{or}\,21\rightarrow 6$.

The $G_{ni, j} \left(\mathbf{x},\mathbf{x}_{\mathrm{s}}, \omega \right)$ is the derivative of the 2D Rayleigh wave green function \citep{snieder19863,aki2002quantitative}, written by,
\begin{equation}
\begin{split}
    G_{ni,j} \left(\mathbf{x},\mathbf{x}_{\mathrm{s}}, \omega \right)   = & A_0  \left\{ {d}_{i,j}^\text{sc} (z_\mathrm{s} ) -\text{i} k_R {e}^\text{sc}_j {d}_i^\text{sc} (z_\mathrm{s} ) \right\} \exp{\left(-\text{i} k_R \mathbf{e}^\text{sc}\cdot \mathbf{x}_{\mathrm{s}}\right)} \frac{\exp \left(\text{i} k_R r\right)}{\sqrt[]{r}}    {p}^\text{sc}_{n} \left(z \right)  \textrm{,}
    \label{eq:green}
    \end{split}
\end{equation}
with
\begin{equation}
    \begin{split}
     A_0  &= \frac{1}{4P_{R}c_R k_R}, \quad  P_R = \frac{1}{2} \rho_0 {c}_{g} \int_{0} ^{\infty } \left[ \lvert U_{R}\left(z\right) \rvert ^2+ \lvert W_{R}\left(z\right) \rvert^2 \right]\mathrm{~d} z, \\
      {p}^\text{sc}_{n} \left(z \right) = & \left[ U_{R}\left(z \right) ,0, W_{R} \left(z \right)  \right] \textrm{,} \quad {d}_{i}^\text{sc}\left(z_\mathrm{s} \right) = \left[-U_{R}(z_\mathrm{s}) , 0 , - W_{R}(z_\mathrm{s})\right]\textrm{,}
    \end{split}
\end{equation}
where $r$ is the propagation distance.   $P_R$ represents a normalised power per unit width in the travelling wave mode.  ${c}_{g}$ is the group velocity of the Rayleigh wave. $\mathbf{e}^\text{sc} =[-1,0,0]$ denotes the propagation direction of the scattered Rayleigh wave.

Substituting Eqs. (\ref{eq:displacement}) and (\ref{eq:green}) into Eq. (\ref{eq:001}) and rearranging the result, we have,
\begin{equation}
{u}^\text{sc}_{n}\left(\mathbf{x}, \omega \right)  = A \left(\omega \right) \frac{\exp\left(\text{i} k_R r\right)}{\sqrt[]{ r}} {p}^\text{sc}_{n} \left(z \right) \textrm{,}
\end{equation}
where $A \left( \omega \right)$ is the far-field amplitude of the backscattered Rayleigh wave, expressed as \citep{li2022theoretical},
 \begin{equation}
\label{eq:009}
\begin{split}
&A\left( \omega \right)  = - A_0  \int    M(\mathbf{x}_{\mathrm{s}})  \exp{\left[\text{i} k_R \left(\mathbf{e}^\text{in}- \mathbf{e}^\text{sc}\right)\cdot \mathbf{x}_{\mathrm{s}} \right]} \mathrm{~d}^2 \mathbf{x}_{\mathrm{s}}  \textrm{,} 
\end{split}
\end{equation}
with
\begin{equation}
\begin{split}
M(\mathbf{x}_{\mathrm{s}}) & =   -k_{R}^2 \Delta C_{i1k1} (\mathbf{x}_{\mathrm{s}})      J_{i k}^{00} +\text{i} k_{R} \Delta C_{i1k3} (\mathbf{x}_{\mathrm{s}}) J_{i k}^{01}
+\text{i} k_{R} \Delta C_{i3k1} (\mathbf{x}_{\mathrm{s}})  J_{i k }^{10}+\Delta C_{i3k3} (\mathbf{x}_{\mathrm{s}})J_{i k}^{11}
\textrm{,}
\\
J_{i k}^{mn}\left(z \right)   &=  \left(-1\right)^{m+n} \sum_{\sigma_1 = L,T}  \sum_{\sigma_2 = L,T} \left(\eta_{\sigma_1}\right)^{m} \left(\eta_{\sigma_2}\right)^{n} 
        w_i^{\sigma_1}
        w_k^{\sigma_2} \exp\left[-\left(\eta_{\sigma_1}+\eta_{\sigma_2}\right)z\right]  \textrm{,}
\end{split}
\end{equation}
where the summation of ($i$, $k$) in the above equations is over 1 and 3. $m, n$ = 0, 1 and no summation convention for repeated $m, n$.
\subsection{\label{sec:2.2} Single scattering of Rayleigh waves from microstructure}
Before  we start with the development of the scattering behaviour of  Rayleigh waves on the polycrystalline material, we make the following assumptions that: (1) the polycrystal's statistics are homogeneous and isotropic; 
(2) there is no orientation correlation between different grains; 
(3) only R-R wave scattering is considered; 
(4) multiple scattering phenomena are neglected, i.e., the reflection from one grain is not affected by other grains; 
(5) total backscattering power is an incoherent sum of the signals scattered by the individual grain in the metal (IS approximation), i.e., the phases of the individual reflection are not correlated.

Based on the above-mentioned assumptions, the backscattered power can be given by \citep{rose1992ultrasonic},
\begin{equation}
\label{eq:011}
\begin{split}
P\left( \omega \right)   =  \left\langle A\left( \omega \right) A^{*}\left( \omega \right)\right\rangle,
\end{split}
\end{equation}
where  $\left< \right>$ denotes the ensemble average. The asterisk represents the conjugate.   $A\left( \omega \right)$ denotes the R-R scattering amplitude in the Born approximation, which is given in Eq. \ref{eq:009}.  Substituting Eq. \ref{eq:009} to Eq. \ref{eq:011}, the total backscattered power can be expressed as,
\begin{equation}
    \begin{split}
     P\left( \omega \right) = A_0^2   \int \int  \Psi \left(\mathbf{x_{\mathrm{s}}}, \mathbf{x}\right)  \exp \left [-\left(\eta_{\sigma_1} +\eta_{\sigma_2}\right)  z_{\mathrm{s}}-\left(\eta_{\sigma_3} +\eta_{\sigma_4}\right) z \right]     \exp{\left[2 \text{i} k_R \left(x_{\mathrm{s}} -x\right)\right]} \mathrm{~d}^2 \mathbf{x}_{\mathrm{s}}
     \mathrm{~d}^2 \mathbf{x}
     \textrm{,}
    \end{split}
\end{equation}
where the summation of each $\sigma_i$ over $L$ and $T$ is implied. In the equation, $\Psi \left (\mathbf{x_{\mathrm{s}}}, \mathbf{x}\right )$ is given by 
\begin{equation}
     \label{eq:013}
    \begin{split}
        \Psi \left (\mathbf{x_{\mathrm{s}}}, \mathbf{x}\right )  & =   \left \langle M (\mathbf{x}_{\mathrm{s}} )  M^{*} (\mathbf{x} ) \right \rangle 
        \\& = k_{R}^4 \left \langle \Delta C_{i 1 k 1}(\mathbf{x}_{\mathrm{s}})\Delta C_{\alpha 1 \gamma 1}(\mathbf{x})\right \rangle \Lambda_{i k  \alpha \gamma}^{0 0 0 0} + \text{i} k_{R}^3 \left \langle \Delta C_{i 1 k 1}(\mathbf{x}_{\mathrm{s}})\Delta C_{\alpha 1 \gamma 3}(\mathbf{x})\right \rangle \Lambda_{i k  \alpha \gamma}^{0 0 0 1}
        \\& + \text{i} k_{R}^3 \left \langle \Delta C_{i 1 k 1}(\mathbf{x}_{\mathrm{s}})\Delta C_{\alpha 3 \gamma 1}(\mathbf{x})\right \rangle \Lambda_{i k  \alpha \gamma}^{0 0 1 0} - k_{R}^2
        \left \langle \Delta C_{i 1 k 1}(\mathbf{x}_{\mathrm{s}}) \Delta C_{\alpha 3 \gamma 3}(\mathbf{x}_{\mathrm{s}}) \right \rangle \Lambda_{i k  \alpha \gamma}^{0 0 1 1}
       \\ &
       -\text{i} k_{R}^3 
       \left \langle \Delta C_{i 1 k 3}(\mathbf{x}_{\mathrm{s}}) \Delta C_{\alpha 1 \gamma 1}(\mathbf{x}) \right \rangle \Lambda_{i k  \alpha \gamma}^{0 1 0 0} 
       + k_{R}^2 
      \left \langle \Delta C_{i 1 k 3}(\mathbf{x}_{\mathrm{s}})\Delta C_{\alpha 1 \gamma 3}(\mathbf{x}) \right \rangle \Lambda_{i k  \alpha \gamma}^{0 1 0 1} 
       \\ &  
       + k_{R}^2 
       \left \langle \Delta C_{i 1 k 3}(\mathbf{x}_{\mathrm{s}}) \Delta C_{\alpha 3 \gamma 1}(\mathbf{x}) \right \rangle \Lambda_{i k  \alpha \gamma}^{0 1 1 0}
       + \text{i} k_{R} 
       \left \langle \Delta C_{i 1 k 3}(\mathbf{x}_{\mathrm{s}})\Delta C_{\alpha 3 \gamma 3}(\mathbf{x})\right \rangle \Lambda_{i k  \alpha \gamma}^{0 1 1 1} 
       \\ &
      - \text{i} k_{R}^3  
      \left \langle \Delta C_{i 3 k 1}(\mathbf{x}_{\mathrm{s}})\Delta C_{\alpha 1 \gamma 1}(\mathbf{x})\right \rangle \Lambda_{i k  \alpha \gamma}^{1 0 0 0}
      + k_{R}^2 
      \left \langle \Delta C_{i 3 k 1}(\mathbf{x}_{\mathrm{s}})\Delta C_{\alpha 1 \gamma 3}(\mathbf{x})\right \rangle \Lambda_{i k  \alpha \gamma}^{1 0 0 1} 
      \\ & 
       + k_{R}^2  
       \left \langle \Delta C_{i 3 k 1}(\mathbf{x}_{\mathrm{s}}) \Delta C_{\alpha 3 \gamma 1}(\mathbf{x})\right \rangle \Lambda_{i k  \alpha \gamma}^{1 0 1 0}
       +\text{i} k_{R} 
       \left \langle \Delta C_{i 3 k 1}(\mathbf{x}_{\mathrm{s}})\Delta C_{\alpha 3 \gamma 3}(\mathbf{x})\right \rangle \Lambda_{i k  \alpha \gamma}^{1 0 1 1} 
        \\ & 
        - k_{R}^2
       \left \langle \Delta C_{i 3 k 3}(\mathbf{x}_{\mathrm{s}})\Delta C_{\alpha 1 \gamma 1}(\mathbf{x})\right \rangle \Lambda_{i k  \alpha \gamma}^{1 1 0 0}
        - \text{i} k_{R}
       \left \langle \Delta C_{i 3 k 3}(\mathbf{x}_{\mathrm{s}})\Delta C_{\alpha 1 \gamma 3}(\mathbf{x})\right \rangle \Lambda_{i k  \alpha \gamma}^{1 1 0 1} 
       \\ &
       -\text{i} k_{R}      
       \left \langle \Delta C_{i 3 k 3}(\mathbf{x}_{\mathrm{s}})\Delta C_{\alpha 3 \gamma 1}(\mathbf{x})\right \rangle \Lambda_{i k  \alpha \gamma}^{1 1 1 0}
       +
       \left \langle \Delta C_{i 3 k 3}(\mathbf{x}_{\mathrm{s}})\Delta C_{\alpha 3 \gamma 3}(\mathbf{x})\right \rangle \Lambda_{i k  \alpha \gamma}^{1 1 1 1}.
    \end{split}
\end{equation}
Meanwhile, $\Lambda^{m n p q}_{i k  \alpha \gamma} $ is defined as,
\begin{equation}
    \begin{split}
        \Lambda_{i k  \alpha \gamma}^{m n p q}  = & J^{m n}_{i k} (J^{ p q}_{\alpha \gamma})^{*} \\  = & \left(-1\right)^{m+n+p+q}  \sum_{\sigma_1 = L,T}\sum_{\sigma_2 = L,T} \sum_{\sigma_3 = L,T} \sum_{\sigma_4 = L,T}     \left(\eta_{\sigma_1}\right)^{m}     \left(\eta_{\sigma_2}\right)^{n} \left(\eta_{\sigma_3}\right)^{p} \left(\eta_{\sigma_4}\right)^{q}  w_i^{\sigma_1} w_k^{\sigma_2} w_\alpha^{\sigma_3*} w_\gamma^{\sigma_4*},
        \end{split}
\end{equation}
for $m, n, p, q$ = 0, 1 and no summation convention for repeated $m, n, p, q$. 

Due to the assumption of statistical homogeneity and macroscopic isotropy of the polycrystalline medium,  the covariance can be factorised into tensorial and spatial parts \citep{stanke1984unified,weaver1990diffusivity,van2018numerical}, 
\begin{equation}
  \left\langle \Delta C_{ijkl}(\mathbf{x}) \Delta C_{\alpha \beta \gamma \delta} (\mathbf{x}_\mathrm{s})\right\rangle =  
  \left\langle \Delta C_{ijkl} \Delta C_{\alpha \beta \gamma \delta}\right\rangle W \left(\mathbf{x}_\mathrm{s} - \mathbf{x} \right),
  \label{eq:015}
\end{equation}
where  $\left\langle \Delta C_{ijkl} \Delta C_{\alpha \beta \gamma \delta}\right\rangle$ is elastic covariance, whose detailed information
 can be found in Ref. \citep{li2022theoretical, weaver1990diffusivity,kube2015acoustic}.  $W\left(\mathbf{x}_\mathrm{s} - \mathbf{x}\right)$ is a geometrical two-point correlation (TPC) function, describing the probability that the two points  $\mathbf{x}$,  $\mathbf{x}_\mathrm{s}$ are in the same grain. 
 For equiaxed grains with a grain size distribution,  Van Pamel \textit{et al.}  \citep{van2018numerical} has proved that  $W(r)$ can be obtained by fitting an exponential series as,
\begin{equation}
\begin{split}
    W\left(\mathbf{x}_\mathrm{s} - \mathbf{x}\right) =   \sum_{\phi=1}^{\Phi} \left[ A_{\phi} \exp \left(-\frac{\left | \mathbf{x}_\mathrm{s} - \mathbf{x} \right |}{a_\phi} \right)\right], \quad \sum_{\phi=1}^{\Phi} A_{\phi} = 1.
    \label{eq:017}
    \end{split}    
\end{equation}
In direct contrast to the conventional single exponential form \citep{stanke1984unified, weaver1990diffusivity}, this generalised TPC function has the advantage of accurately representing the actual TPC statistics of experimental \citep{man2006geometric} and numerical samples \citep{van2018numerical}. The equivalent grain coefficients $a_\phi$ and $A_\phi$ are determined by best fitting the actual TPC data of the polycrystal, which is discussed in Sec.\ref{sec:3}. 

Substituting Eqs. \ref{eq:015} and \ref{eq:017}  to Eq. \ref{eq:013}, we can get,
\begin{equation}
    \begin{split}
     P\left( \omega \right)  =   A_0^2  \Psi_{0}    \sum_{\phi = 1}^{\Phi} \Biggl \{ A_{\phi}   \int \int 
      &  \Big [  \exp \left(-\frac{\left | \mathbf{x}_\mathrm{s} - \mathbf{x} \right |}{a_\phi}\right)  
 \exp \left[-\left( \eta_{\sigma_1} +\eta_{\sigma_2} \right)  z_{\mathrm{s}}-\left(\eta_{\sigma_3} +\eta_{\sigma_4}) z \right) \right]  \\ & \times   \exp{\left[2 \text{i} k_R \left(x_{\mathrm{s}} -x \right)\right]} \mathrm{~d}^2 \mathbf{x}_{\mathrm{s}}
     \mathrm{~d}^2 \mathbf{x} \Big ] \Biggl \}
     \textrm{,}
      \end{split}
\end{equation}
with
\begin{equation}
\begin{split}
\Psi_{0}  = & k_{R}^4
\left \langle \Delta C_{i 1 k 1}\Delta C_{\alpha 1 \gamma 1}\right \rangle \Lambda_{i k  \alpha \gamma}^{0 0 0 0} 
+ \text{i} k_{R}^3
\left \langle \Delta C_{i 1 k 1}\Delta C_{\alpha 1 \gamma 3}\right \rangle \Lambda_{i k  \alpha \gamma}^{0 0 0 1}  
+ \text{i} k_{R}^3
\left \langle \Delta C_{i 1 k 1}\Delta C_{\alpha 3 \gamma 1}\right \rangle \Lambda_{i k  \alpha \gamma}^{0 0 1 0} 
\\&
- k_{R}^2
\left \langle \Delta C_{i 1 k 1} \Delta C_{\alpha 3 \gamma 3} \right \rangle \Lambda_{i k  \alpha \gamma}^{0 0 1 1}
- \text{i} k_{R}^3 
\left \langle \Delta C_{i 1 k 3} \Delta C_{\alpha 1 \gamma 1} \right \rangle \Lambda_{i k  \alpha \gamma}^{0 1 0 0} 
+ k_{R}^2 
\left \langle \Delta C_{i 1 k 3}\Delta C_{\alpha 1 \gamma 3} \right \rangle \Lambda_{i k  \alpha \gamma}^{0 1 0 1} 
\\&  
+ k_{R}^2
\left \langle \Delta C_{i 1 k 3} \Delta C_{\alpha 3 \gamma 1} \right \rangle \Lambda_{i k  \alpha \gamma}^{0 1 1 0}
+ \text{i} k_{R} 
\left \langle \Delta C_{i 1 k 3}\Delta C_{\alpha 3 \gamma 3}\right \rangle \Lambda_{i k  \alpha \gamma}^{0 1 1 1} 
- \text{i} k_{R}^3
\left \langle \Delta C_{i 3 k 1}\Delta C_{\alpha 1 \gamma 1}\right \rangle \Lambda_{i k  \alpha \gamma}^{1 0 0 0}
\\ &
+ k_{R}^2
\left \langle \Delta C_{i 3 k 1}\Delta C_{\alpha 1 \gamma 3}\right \rangle \Lambda_{i k  \alpha \gamma}^{1 0 0 1} 
+ k_{R}^2
\left \langle \Delta C_{i 3 k 1} \Delta C_{\alpha 3 \gamma 1}\right \rangle \Lambda_{i k  \alpha \gamma}^{1 0 1 0} 
+\text{i} k_{R}
\left \langle \Delta C_{i 3 k 1}\Delta C_{\alpha 3 \gamma 3}\right \rangle \Lambda_{i k  \alpha \gamma}^{1 0 1 1} 
\\ & 
- k_{R}^2 
\left \langle \Delta C_{i 3 k 3}\Delta C_{\alpha 1 \gamma 1}\right \rangle \Lambda_{i k  \alpha \gamma}^{1 1 0 0}
- \text{i} k_{R}
\left \langle \Delta C_{i 3 k 3}\Delta C_{\alpha 1 \gamma 3}\right \rangle \Lambda_{i k  \alpha \gamma}^{1 1 0 1} 
-\text{i} k_{R}
\left \langle \Delta C_{i 3 k 3}\Delta C_{\alpha 3 \gamma 1}\right \rangle \Lambda_{i k  \alpha \gamma}^{1 1 1 0}
\\&
+\left \langle \Delta C_{i 3 k 3}\Delta C_{\alpha 3 \gamma 3}\right \rangle \Lambda_{i k  \alpha \gamma}^{1 1 1 1}.
    \end{split}
\end{equation}
By applying the following change of variables, 
\begin{equation}
   \mathbf{\boldsymbol{\tau}}   = \left(\mathbf{x}_{\mathrm{s}}+\mathbf{x}\right) \big / 2, \quad \mathbf{r} = \mathbf{x}_\mathrm{s}-\mathbf{x},
\end{equation}
with the limit of (the detailed information for the transformation of the integral area in the Appendix),
\begin{equation}
 0<\tau_z<+\infty, \  -\infty<\tau_x<+\infty, \ -\infty<r_z<+\infty, \ -\infty<r_x<+\infty . 
\end{equation}

Therefore, the following equation is straightforward,
\begin{equation}
    \begin{split}
     P\left( \omega \right)  =    A_{0}^2 \Psi_{0}     \sum_{\phi = 1}^{\Phi} \left \{ A_{\phi} \int  \exp \left(-\eta_M \tau_z\right) \mathrm{~d}^2  \mathbf{\boldsymbol{\tau}}   \int \exp \left(- {r} \big / {a_\phi}+2 \text{i} k_R r_x+\eta_N r_z\right )  \mathrm{~d}^2  \mathbf{r} \right \}  \textrm{,}
    \end{split}
\end{equation}
where  $\eta_M = \eta_{\sigma_1} +\eta_{\sigma_2} +\eta_{\sigma_3} +\eta_{\sigma_4}$ and $\eta_N = \left(-\eta_{\sigma_1} -\eta_{\sigma_2}+\eta_{\sigma_3} +\eta_{\sigma_4}\right)\big /{2}$.

We want to emphasise that, unlike the uniform bulk wave, there is an exponential energy decay for Rayleigh waves in the $z-$ displacement (thickness direction). Thus, instead of estimating the power per unit area, \textit{the backscattered power $p\left( \omega \right)$}, from the grains in the area where is the unit length multiplied by the infinite depth, should be used to assess the backscattering behaviour of Rayleigh waves propagating on the material surface. Based on this, we make a calculation for the integral, which can be followed, 
\begin{equation}
    \begin{split}
     p\left( \omega \right)  =    A_0^2  \Psi_{0}    \sum_{\phi = 1}^{\Phi} \Biggl \{ A_{\phi} 
 \int_{0}^{\infty}   \exp \left(-\eta_M \tau_z\right) \mathrm{~d}  {\tau_z}  \int \exp \left(- {r}\big /{a_\phi}+2 \text{i} k_R r_x+\eta_N r_z\right ) \mathrm{~d}^2   \mathbf{r} \Biggl \} \textrm{.}
     \end{split}
\end{equation}
After further manipulation, the final equation can be written as,
\begin{equation}
    \begin{split}
     p\left( \omega \right)  =     {A_0^2  \Psi_{0} }  \sum_{\phi = 1}^{\Phi} \left \{ A_{\phi}  \frac{2 \pi a_\phi^2}{\eta_M \left[a_\phi^2 \left(4 k_R^2 - \eta_N^2 \right) + 1\right]^{3/2}} \right \}  \textrm{.}
     \label{eq:025}
       \end{split}
\end{equation}

The theoretical prediction of the rms backscattering amplitude of single scattering for multiple grains can be given by, 
\begin{equation}
    A_{\mathrm{rms}}\left( \omega \right) = \sqrt{{p\left( \omega \right)} \big / {n_g}},
    \label{eq:026}
\end{equation}
where $n_g$ is grain density. Assuming that $N$ is the number of grains in the active volume of a sample.  $\Omega$ is the active space of the sample.  $n_g  = N /  \Omega $. $A_{\mathrm{rms}}\left( \omega \right)$ is the backscattering amplitude of Rayleigh waves for a single scattering of multiple grains with the area (unit length $\times$ infinite depth), which is the most important result in this article.

\subsection{\label{sec:2.3} Grain noise of 
Rayleigh waves in a polycrystalline material}

Now, we consider a 2D case where one point on the upper surface ($x-$ surface) transmits a plane Rayleigh wave and receive the backscattered and reflected signals. Then, the normalised backward grain noise $N_{\mathrm{rms}}$ in the area with unit length $\times$ infinite depth can be defined by the rms of backward amplitude from each grain, given as an approximate expression for the dimensionless ratio  \citep{margetan1993modeling},
\begin{equation}
   N_{\mathrm{rms}}\left( \omega \right)\equiv \sqrt{\left<\big \lvert \Gamma_{\mathrm{noise}}\left( \omega \right) \big \rvert^2\right> \bigg / \big \lvert \Gamma_{\mathrm{ref}}\left( \omega \right) \big \rvert ^2},
   \label{eq:027}
\end{equation}
where $\Gamma_{\mathrm{ref}}\left( \omega \right)$ is the Fourier transform of the reference signal at angular frequency $\omega = 2 \pi f$.  $\Gamma_{\mathrm{noise}}\left( \omega \right)$ denotes the Fourier transform of the grain noise signal on the finite time interval indicated in the area with unit length $\times$ infinite depth, which is understood to be located away from the reference echoes. 

In addition to the five assumptions we mentioned in Sec. \ref{sec:2.2}, here we apply two more restrictions: (6) No attenuation is considered, which is the direct consequence of assumption (4) above which stipulates that the reflection from one grain is not affected by other grains; (7) The time window of interest is long enough to enclose the time-domain echoes produced by the backscattering of sound by all grains in some regions of the specimen. 

As we mentioned in assumption (5), the IS approximation states that for an incoherent summation of signals, the power of the sum equals the sum of the powers of the contributing signals. Thus,
\begin{equation}
  {\left< \big \lvert \Gamma_{\mathrm{noise}}\left( \omega \right) \big \rvert^2\right>} = \left<\sum \big \lvert\Gamma_i\left( \omega \right)^2 \big \rvert\right>,
      \label{eq:028}
\end{equation}
considering an echo associated with Rayleigh waves which travel directly from the transmitter to grain at position $\mathbf{x}_{i} (x_i,z_i)$ and then directly back to the receiver, the discrete Fourier transform component of this echo at frequency $f$ may be approximated by \citep{thompson1983model},
\begin{equation}
     \Gamma_i(\omega,x_{i},z_{i}) = {A_i\left( \omega \right) }\Gamma_{\mathrm{ref}}\left( \omega \right),
    \label{eq:029}
\end{equation}
where $A_i\left( \omega \right)$ is the scattering amplitude from this grain, given by Eq. \ref{eq:009}.

Furthermore, replacing the sum over grains with an integral over the unit volume of the material, Eq. \ref{eq:028} can be given as, 
\begin{equation}
  {\left<\big \lvert \Gamma_{\mathrm{noise}}\left( \omega \right) \big \rvert^2\right>} = \left<\sum \big \lvert A_i\left( \omega \right)^2 \big  \rvert \right> \big \lvert \Gamma_{\mathrm{ref}}\left( \omega \right) \big \rvert ^2 =  p\left( \omega \right) \big \lvert \Gamma_{\mathrm{ref}}\left( \omega \right) \big \rvert^2,
  \label{eq:030}
\end{equation}
where $p\left( \omega \right)$ is given by Eq. \ref{eq:025}. Substitution of Eq. \ref{eq:030}  in Eq. \ref{eq:027}, the normalised grain noise in the area with unit length $\times$ infinite depth  can be further obtained as,
\begin{equation}
        N_{\mathrm{rms}}\left( \omega \right) = \sqrt{p\left( \omega \right)} ={\sqrt{n_g}}{A_{\mathrm{rms}}\left( \omega \right)},
    \label{eq:031}
\end{equation}
where $A_{\mathrm{rms}}\left( \omega \right)$ is written as Eq. \ref{eq:026}. From Eq. \ref{eq:031}, we can figure out that $p\left( \omega \right)$ is the square of FOM to represent the measurement of the inherent grain noise of the specimen of Rayleigh wave and can be also expressed as the backscattering coefficient, which can be written as $\text{FOM}^2 \equiv p\left( \omega \right) $.

We have now completed our theoretical developments for the quantitative predictions of the rms of the backscattering Rayleigh wave in (Eqs. \ref{eq:026} and \ref{eq:031}). For the verification of these predictions, we computationally simulate the rms backscattering amplitude of single grains with random shapes and orientations and grain noise scattered by the polycrystalline material in the following section.

\section{\label{sec:3}Finite element method}

The capability of 2D FE to model the bulk wave scattering in polycrystals has been well proved in recent FE modelling papers  \citep{van2018numerical,van2015finite,van2017finite,huang2021longitudinal}. Therefore, numerical validations are used in this section to verify the theoretical model. The FE method for simulating Rayleigh waves flaw scattering was implemented in our prior work  \citep{li2022theoretical}. A brief overview of the FE method is given below and several previous aspects are emphasised here.

\begin{figure}[htp]
\centering
\includegraphics[width=0.7\textwidth]{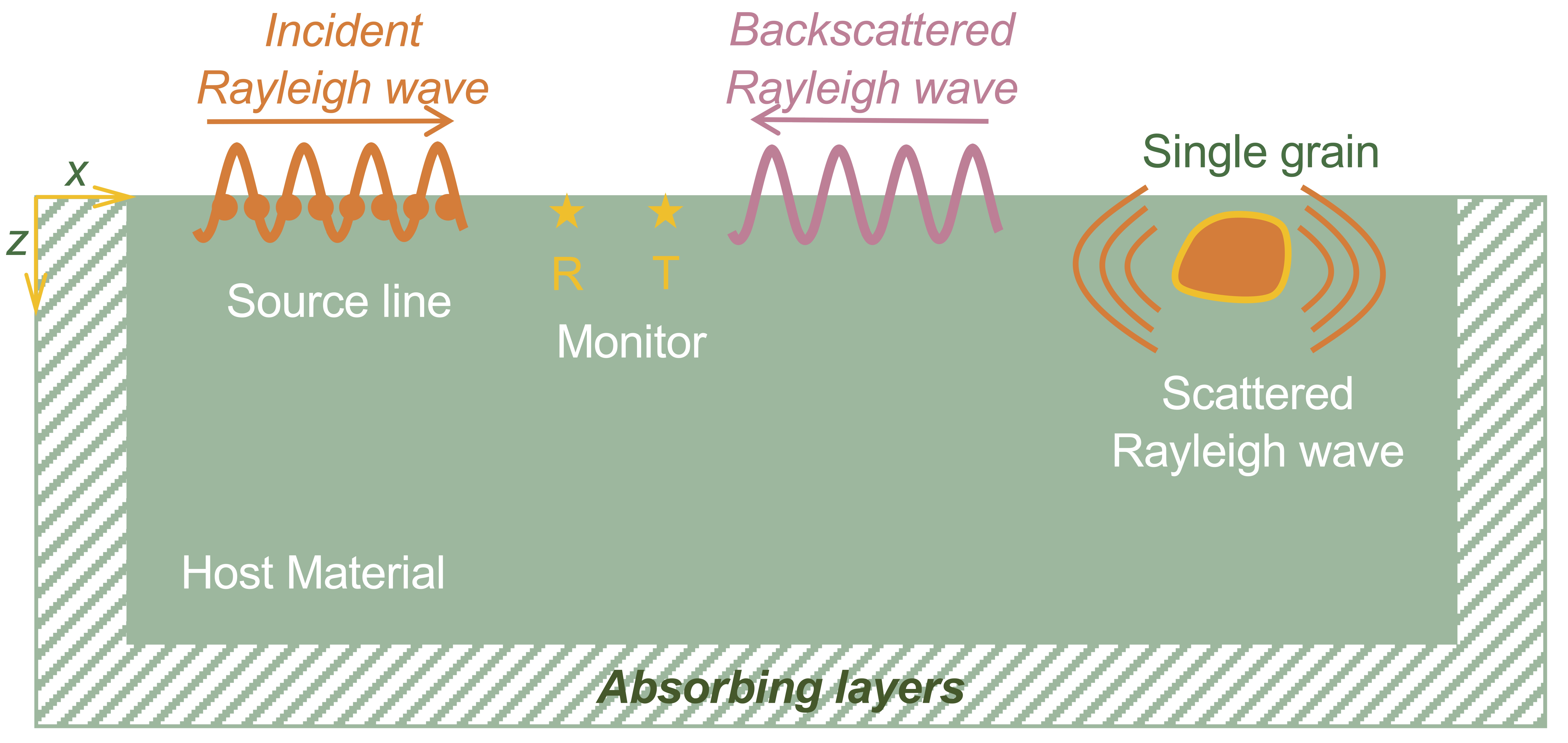}
\caption{\label{fig:1} Example FE modelling of Rayleigh wave scattering by single grain.}
\end{figure}

As schematically shown in Fig. \ref{fig:1}, the 2D FE model is based on the $x-z$ plane. The numerical polycrystalline models used here are constructed in the Neper program  \citep{quey2011large} with the Poisson Voronoi tessellations (PVT). The PVT creates uniformly random seeds in the model space of a polycrystal, with each seed being enclosed by a grain within which all points are closer to the enclosed seed than to any other  \citep{huang2022finite}. The grains are statistically equiaxed because of the procedure of randomly placing Voronoi seeds \citep{van2018numerical,van2015finite}. 

\begin{figure}[htp]
\centering
\includegraphics[width=0.8\textwidth]{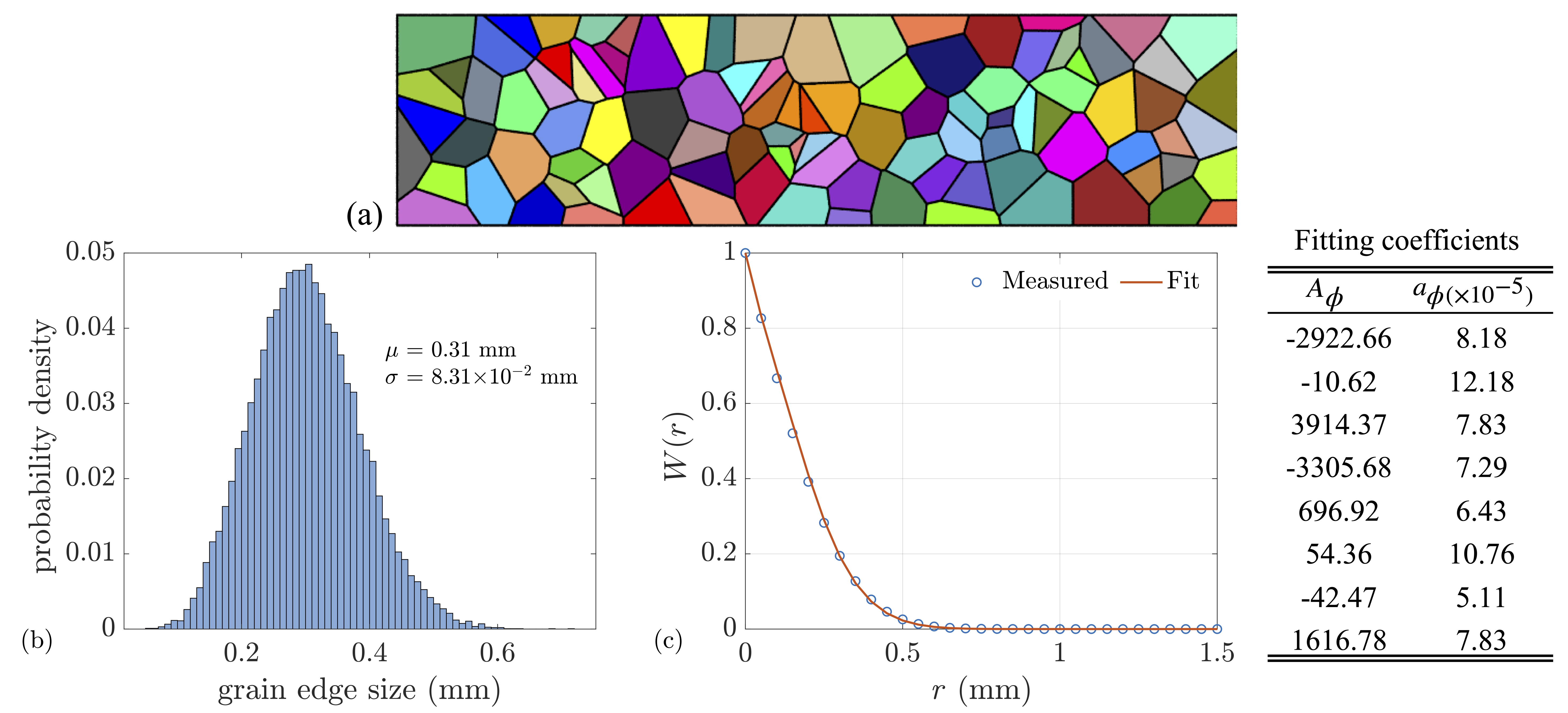}
\caption{\label{fig:2} TPC statistics for the polycrystalline microstructures with Poisson Voronoi tessellations. (a) Polycrystalline microstructure with a dimension of 140 $\times$ 14 mm and 20 000 grains. (b) the grain size distribution is represented by the probability density of grain size. (c) TPC statistics with the measured and mathematical fit. The fitting coefficients are also shown.}
\end{figure}

Taking the model n20000 in Table \ref{tab:1} with the PVT microstructure as an example:  its dimensions $d_x \times d_z$ are 140 mm $\times$ 14 mm, the averaged grain edge size $\Bar{d}$, defined as the square root of the space area divided by the grain number \citep{van2015finite,huang2021longitudinal}, is 0.31 mm, and the polycrystal microstructure is displayed in Fig. \ref{fig:2}(a). The grain edge sizes, defined as the square root of each grain area, of the PVT grains are normally distributed \citep{ryzy2018influence} and shown in Fig. \ref{fig:2}(b). The mean grain size is 0.31 mm, with a standard deviation of 8.31 $\times 10^{-2}$ mm. The TPC statistics $W(r)$ are numerically measured from the generated polycrystal models and the resulting data points are indicated in Fig. \ref{fig:2}(c).  To incorporate the measured statistics into the theoretical models, they are fitted into a generalised TPC function (Eq. \ref{eq:017}), which is displayed in Fig. \ref{fig:2}(c) as the solid curve. The fitted TPC function is treated as a scalar function and the detailed information related to $a_\phi$ and $A_\phi$ can be found in Fig.\ref{fig:2}(c). We note that the fitting numbers  $a_\phi$ and $A_\phi$ are obtained by scaling the fitting parameters mentioned in prior work \citep{huang2020maximizing}. Besides, some other tessellations have been widely used to generate microstructures, such as centroidal Voronoi tessellation \citep{ryzy2018influence, huang2022finite} and Laguerre tessellation \citep{bourne2020laguerre, quey2018optimal}, which are not shown in this paper because of the high computational requirements. Meanwhile, we want to underscore that the theoretical model is reliant on the TPC statistics, which means the theoretical model can be applied with a well-defined TPC function regardless of the type of tessellation model employed.

Structured meshes, which have been shown to perform well with sufficiently fine discretization in modelling a grained material \citep{van2015finite}, continue to be used here. The mesh size used for each model has met the two requirements for obtaining accurate simulation results: (1) at least ten elements per wavelength \citep{huang2020maximizing}; (2) at least ten elements per averaged grain size \citep{van2015finite}. The elements on the bottom, left, and right sides of the model are used to define absorbing boundary conditions. The thickness of each absorbing boundary region in the boundary normal direction is chosen to be three times the wavelength of the Rayleigh wave in the host material  \citep{rajagopal2012use}. 

The desired Rayleigh wave is generated by applying two sinusoidal time-domain signals of 90$^\circ$ phase shift to multiple source nodes located on the top surface of the model (yellow points in Fig. \ref{fig:1}). The size of the source is set to be equal to three times of centre-frequency wavelengths of the simulated Rayleigh wave, and each source node is assigned a unique amplitude, following Eq. (17) in Sarris et al. \citep{sarris2021attenuation}. The simulation is solved using the GPU-accelerated Pogo program  \citep{huthwaite2014accelerated} with an explicit time-stepping scheme. A relatively large time step of $\Delta t = 0.9 h /c_L$, satisfying the Courant-Friedrichs-Lewy condition  \citep{Gnedin2018}, is used to minimise numerical error  \citep{huang2020maximizing}. The models generated for this work are summarised in Table \ref{tab:1}.

\begin{table}[t]
\centering
\caption{\label{tab:1} Models used in the simulation. Dimensions $d_x \times d_z$  (mm $\times$ mm), number of grains $N$, average grain size $\Bar{d}$ (mm), mesh size $h$ (mm), degree of freedom (d.o.f), centre frequency of FE modelling $f_c$ (MHz),}
\begin{adjustbox}{width= \textwidth}
\begin{tabular}{ccccccccc}
\hline\hline
\multirow{1}{*}{Material} &
\multirow{1}{*}{Model} &
\multirow{1}{*}{ $f_c$ (MHz)}  &
\multirow{1}{*}{$d_x \times d_z$ (mm $\times$ mm)}  &
\multirow{1}{*}{$N$}&
\multirow{1}{*}{$\bar{d}$ (mm)}&
\multirow{1}{*}{$h$ (mm)}  & \multirow{1}{*}{d.o.f} 
\\ \hline
Aluminium, A=1.5 & n80000         &  1                                        & 280  $\times $ 28        &   80 000                     &   0.31        &   40 $\times$ 10$^{-3}$  &  9.8 $\times$ 10$^6$
\\ \hline
Aluminium, A=1.5 & n20000            &   2                                     & 140  $\times $ 14                          & 20 000      &  0.31      &   24 $\times$ 10$^{-3}$  &  6.8 $\times$ 10$^6$
\\
Inconel, Lithium & m20000             &   2                              & 160  $\times $ 16                   & 20 000           &   0.36           &   40 $\times$ 10$^{-3}$  &   3.2 $\times$ 10$^6$
\\ \hline
Aluminium, A=1.5 & n5000          &   4                                 & 70  $\times $ 7                   &  5 000           &   0.31                &   12 $\times$ 10$^{-3}$  &     6.8 $\times$ 10$^6$
\\ 
Inconel, Lithium & m5000               &   4                            & 80  $\times $ 8                   &  5 000           &   0.36           &   24 $\times$ 10$^{-3}$  &   2.2 $\times$ 10$^6$
\\ \hline\hline
\end{tabular}
\end{adjustbox}
\end{table}

\section{\label{sec:4}Results and discussions}

\subsection{\label{sec:4.1}rms  backscattering of single grains with random shapes and orientations}

First, rms backscattering of single grains with random shapes and orientations is investigated. We know that the theoretical model is developed with independent scattering approximation. Therefore, the FE model, simulating the backscattering of single grains embedded in a homogeneous background material, can avoid the effect of multiple scattering, which makes the FE results represent the theoretical result better.

\begin{figure}[htp]
\centering
\includegraphics[width=0.7\textwidth]{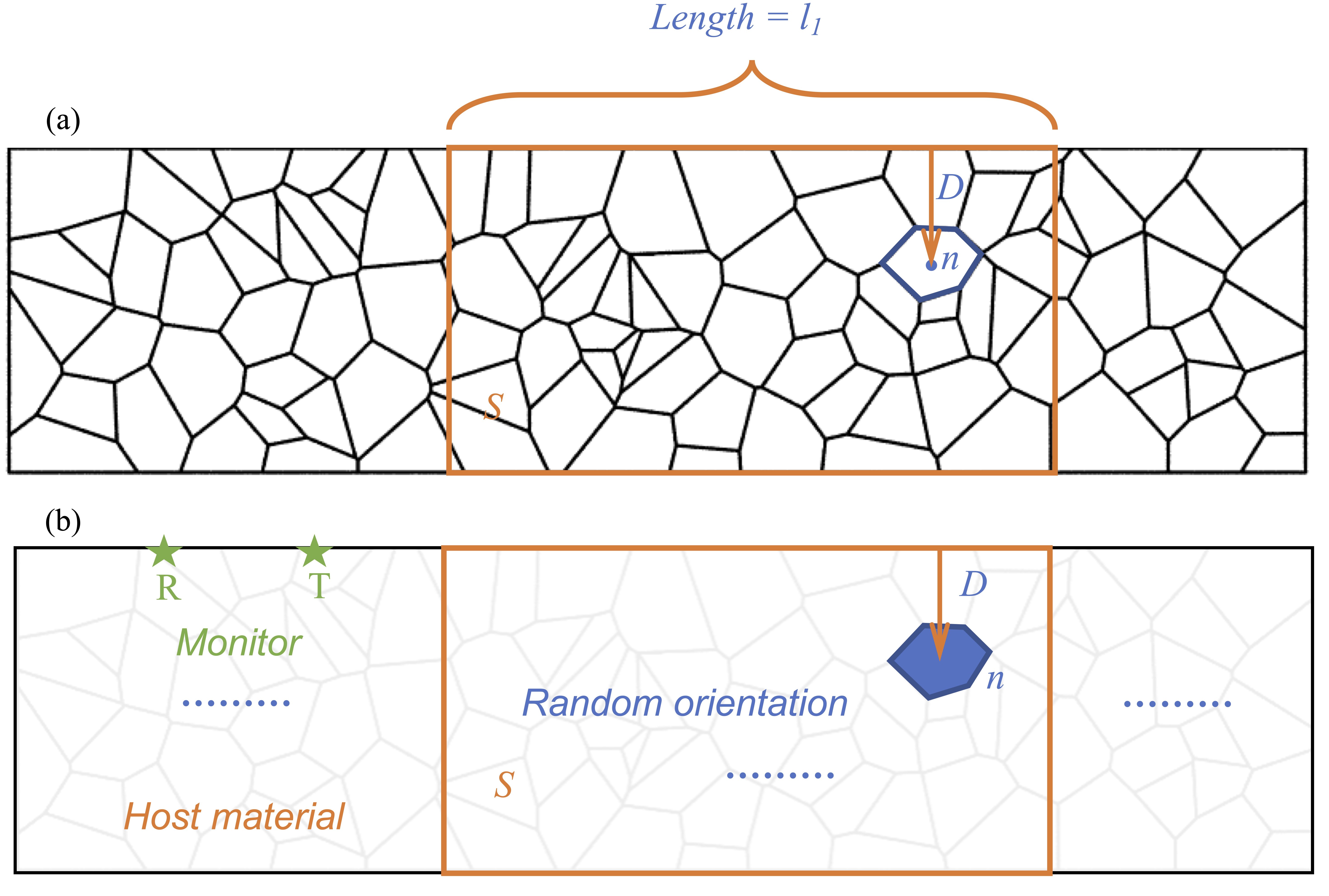}
\caption{\label{fig:3} The schematic of finite element models for the single grains with random shapes and material properties. (a) is the schematic of multiple grain aggregate; (b) a simulation diagram for the single scattering of the Rayleigh wave. The grains in the $S$ area are all used to perform the simulations. 
Each simulation employs only one grain within the $S$ area (thus forming a single scatterer case) with a random orientation while preserving the original location and shape information. The host material is considered homogeneous and isotropic. The transmitted Rayleigh wave and the received backscattered wave are monitored at points `T' and `R', respectively.  The length of $S$ area equals $l_1$. }
\end{figure}

To predict the rms backscattering amplitude of single grain with random shapes and orientations, a number of grains are needed. As we mentioned in Sec. \ref{sec:3}, Neper program generates an aggregate of grains (`multi-grain' models) with each having its own geometry. The geometrical schematic of multiple grain aggregate is shown in Fig. \ref{fig:3}(a).  In order to perform simulations with only single R-R scattering while avoiding the effect of multiple scattering (to be discussed in Sec. \ref{sec:4.2}), we make the following simplification to the simulation model: instead of Rayleigh waves propagating on the whole polycrystalline model directly, the simulation process takes out an individual grain in the $S$ from the polycrystal, and applies a random orientation to it, then embeds it in an isotropic host material, and then simulates the Rayleigh wave propagation in this model with the single grain regarded as a sole scatterer, as shown in Fig. \ref{fig:3}(b). We repeat the process within the $S$ area until enough grains are used to make sure that the grain distribution is statistically uniform. 

Over the course of the FE solution, the $z$ - displacement of the generated incident wave is monitored at a transmitting node (point T in Fig. \ref{fig:3} ), while that of the backscattered wave is recorded at a receiving node (point R). We emphasise that the transmitting and receiving nodes are placed respectively far away from the source nodes and the grain, in order for the former to monitor the well-formed incident wave and for the latter to record solely the scattered Rayleigh wave in the far field. In addition, a reference signal is obtained at the receiving point using an identical but grain-free FE model, and the reference signal is subtracted from the relatively small raw signal to minimise the influence of numerical error. 

\begin{figure}[htp]
\centering
\includegraphics[width=1\textwidth]{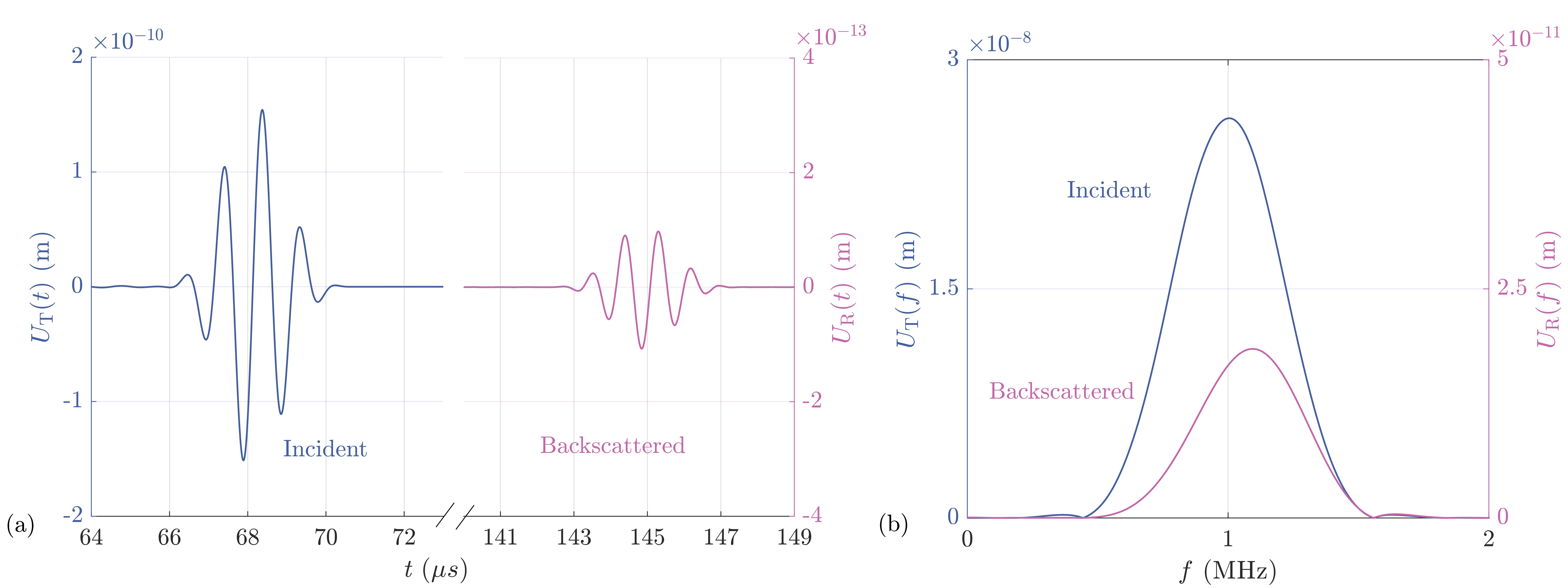}
\caption{\label{fig:4} (a) The $z$- displacements of the incident Rayleigh wave and the backscattered Rayleigh wave in the time domain. (b) the related amplitude spectra in the frequency domain.}
\end{figure}

The backscattering amplitude $ A^{\mathrm{FE}}\left(f \right) $ from single grains with different shapes and random orientations will be measured to calculate the incoherent rms averaging $A_{\mathrm{rms}}^{\mathrm{FE}}\left( f \right)$. Here, a brief overview of the measurement of the backscattering amplitude and rms averaging is introduced. Figure \ref{fig:4}(a) and (b) present an example to illustrate the $z-$ displacement monitored by the transmitter and receiver in the time domain and frequency domain for the ${i}$-th grain, respectively. The signal $U^{{i}}_{\mathrm{T}}\left( t \right)$ at the transmitting node and the corrected signal $U^{{i}}_{\mathrm{R}}\left( t \right)$ at the receiving node are Fourier transformed into the frequency domain to obtain the spectra $U^{{i}}_{\mathrm{T}}\left( f \right)$ and $U^{{i}}_{\mathrm{R}}\left( f \right)$. The frequency-dependent amplitude of the backscattered Rayleigh wave is then calculated by

\begin{equation}
    A^{i}\left( f \right) = U_{\mathrm{R}}^{i}\left( f \right) \big / U_{\mathrm{T}}^{i}\left( f \right)  \quad \text{and} \quad A_{\mathrm{rms}}^{\mathrm{FE}}\left( f \right) = \textbf{RMS}\left[A^{i}\left( f \right)\right]\big /\sqrt{l_1} \textrm{,} 
    \label{eq:Arms}
\end{equation}
where $l_1$ is the the length of $S$ area. $A_{\mathrm{rms}}^{\mathrm{FE}}\left( f \right)$ will be used to evaluate the theoretical model result, $A_{\mathrm{rms}}\left( \omega \right)$, in Sec. \ref{sec:2.2}. 

\begin{figure}[htp]
\centering
\includegraphics[width=1\textwidth]{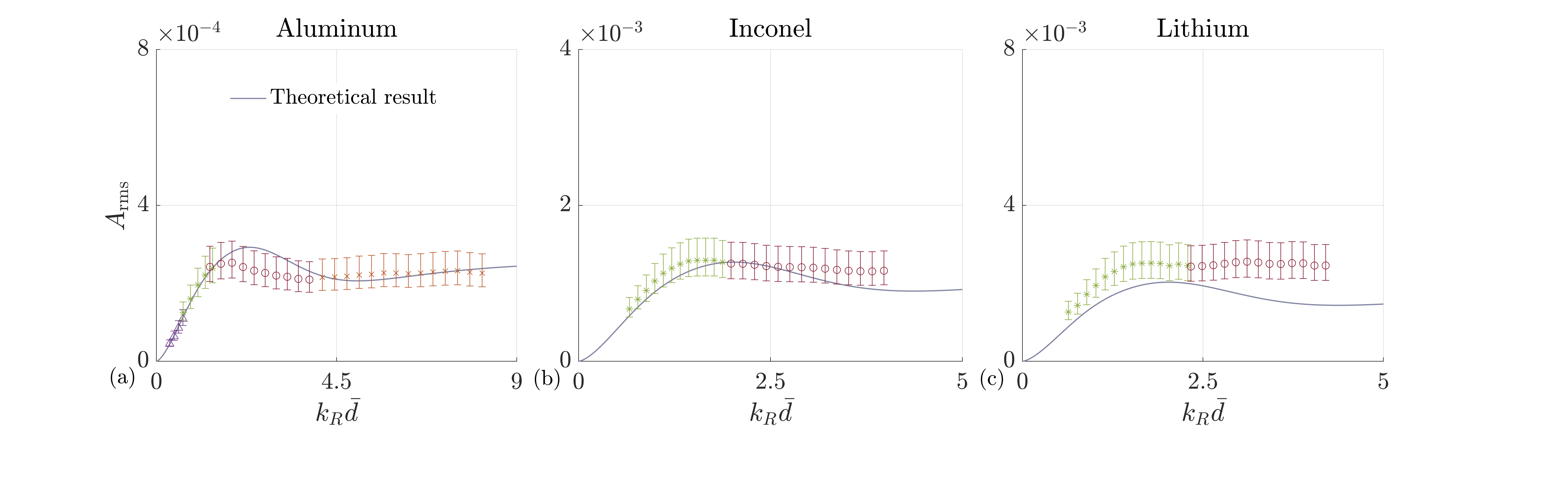}
\caption{\label{fig:5} Comparison of FE and theoretical rms backscattering amplitudes of single random-shaped grains with a cubic material whose anisotropy factors are (a) 1.24 (aluminium), (b) 2.83 (Inconel) and (c) 9.14 (lithium), respectively. The error bars show the $99.73\%$ confidence interval ($3 \sigma$ rule)  \citep{spiegel2013probability} for the FE points.}
\end{figure}

Figure \ref{fig:5}(a) shows the comparison of the FE and theoretical predictions with aluminium material whose detailed information is listed in Table \ref{tab:2}. The $x$ axis is the product of the wave number $k_R$ and  the average linear dimension of the grains in the `multi-grain' models, denoted by $\Bar{d}$ (the detailed information about $d$ is in Table \ref{tab:1}). The FE predicted RMS backscattering amplitude (calculated by Eq. \ref{eq:Arms}) is plotted as the coloured dots and the error bars show the $99.73\%$ confidence interval ($3 \sigma$ rule)  \citep{spiegel2013probability} for the FE points, demonstrating the variation across the realisations with different crystallographic orientations. Four centre frequencies, 1, 2, 4, and 8 MHz, are used in the FE simulations to cover a large range of $k_R\Bar{d}$. The theoretical prediction calculated using Eq. \ref{eq:026} is plotted as the grey curve. The good agreement between the FE results and theoretical prediction shows that the theoretical model is correct and therefore it has a strong potential to predict grain noise, which will be discussed later in Sec. \ref{sec:4.2}.

Furthermore, to observe how the agreement changes with anisotropy, the comparisons for the Inconel and lithium materials which have anisotropy indices of 2.83 and 9.14 are also performed, as shown in Figs. \ref{fig:5}(b) and (c). It is clearly demonstrated that the agreement between the theoretical and FE results decreases as the anisotropy index increases. Such results are reasonable because the theoretical model is developed based on the Born approximation which is expected to gradually fail with the increase of scattering intensity.

In addition, it is shown in Fig. \ref{fig:5} that theoretical values are always smaller than FE results in the large $k_R \Bar{d}$ region, which means that the Born approximation would always result in an underestimation for the $A_\mathrm{rms}$ values in the large $k_R \Bar{d}$ region. Meanwhile, we note that all the simulation results on high regions seem flat with fewer fluctuations than theoretical predictions. A possible reason for the difference is that simulation numbers are insufficient; however, a compromise must be made between accuracy and computational time.

\begin{table}[t]
\centering
\caption{\label{tab:2} Properties of polycrystalline material. Density $\rho$ (kg/m$^3$), equivalent anisotropy index \citep{sha2018universal,huang2022appraising} $A^{eq}$ (note that $A^{eq}$ equals to the Zener anisotropy index $A$ for cubic materials considered here), elastic constants $c_{ij}$ (GPa), and Voigt-averaged $c_{11}^0$ and $c_{44}^0$ (GPa).}
\begin{adjustbox}{width= \textwidth}
\begin{tabular}{llcccccccccccccc}
\hline\hline
 Material \citep{huang2022finite} & \multirow{1}{*}{$\rho$ (kg/m$^3$)}  & \multirow{1}{*}{$A^{eq}$} & \multirow{1}{*}{$c_{11}$ (GPa)}  & \multirow{1}{*}{$c_{12}$ (GPa)}  & \multirow{1}{*}{$c_{44}$ (GPa)}  &
\multirow{1}{*}{$c_{11}^0$ (GPa)} &
\multirow{1}{*}{$c_{44}^0$ (GPa)} &\\ \hline
Aluminium & 2700 & 1.24 & 106.7 & 60.4  & 28.3 & 110.8 & 26.2 \\ 
A = 1.5  & 8000 & 1.52 & 262.1 & 136.5  & 95.3 & 288.1  &  82.3 \\ 
Inconel  & 8260 & 2.83 & 234.6 & 145.4  & 126.2 & 299.9 & 96.6 \\ 
Lithium  & 534 & 9.14 & 13.4 & 11.3  & 9.6 & 20.4 & 6.18 \\ 
                        \hline \hline
\end{tabular}
\end{adjustbox}
\end{table}

\begin{figure}[htp]
\centering
\includegraphics[width=0.5\textwidth]{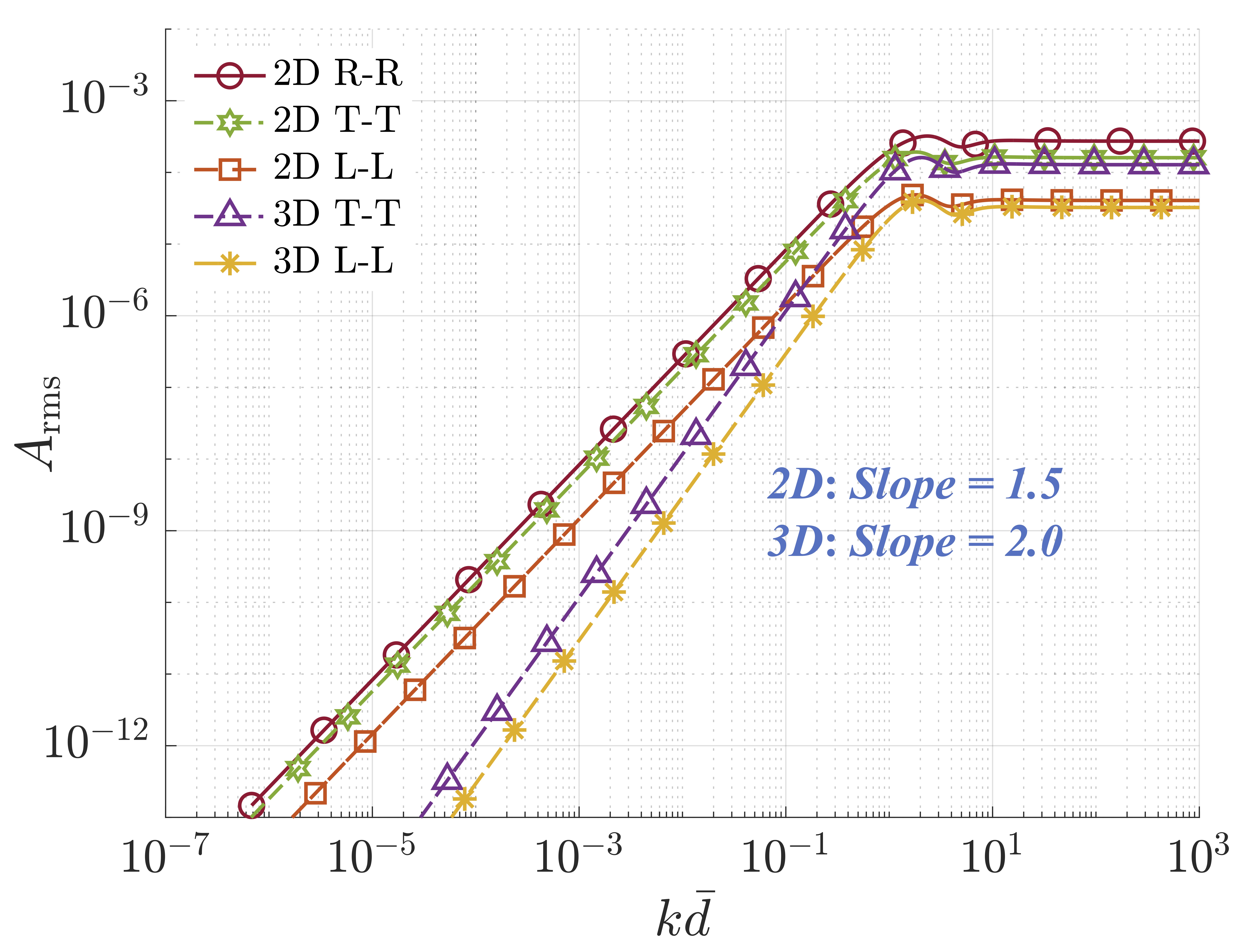}
\caption{\label{fig:6} The relationship between the backscattering amplitude and frequency with different theoretical backscattering models, including 2D R-R scattering, 2D T-T scattering, 2D L-L scattering, 3D T-T scattering and 3D L-L scattering. Aluminium material is used.}
\end{figure}

Now, we discuss the quantitative connection between the backscattering amplitude of Rayleigh waves with frequency.  Figure \ref{fig:6} demonstrates the logarithm of the rms backscattering amplitude $A_{\mathrm{rms}}$ versus the logarithm of normalised frequency $k_R \Bar{d}$ for the aluminium material. Meanwhile, the comparison between the 2D R-R scattering, 2D bulk wave scattering\citep{bai2018finite,ghoshal2009diffuse} and 3D bulk wave scattering \cite{rose1992ultrasonic,hu2017transverse,huang2021transverse,Guo2003effects,gubernatis1977formal} including longitudinal-to-longitudinal (L-L) scattering and transverse-to-transverse (T-T) scattering,  is also performed.   It can be clearly seen that the quantitative relationship  between the backscattering amplitude and the normalised frequency for 2D Rayleigh waves and 2D bulk waves is similar. For the Rayleigh waves, the backscattering amplitude is proportional to one and a half power of frequency for wavelengths much larger than the average grain size or comparable to the size of the average grain ($k_R \Bar{d} < 1$). For shorter wavelengths ($k_R \Bar{d} > 10^{1}$), the backscattering amplitude saturates and becomes independent of frequency. Moreover, the backscattering amplitude of Rayleigh waves is obviously larger than that of bulk waves(2D scattering and 3D scattering). It implies that there is a stronger scattering for 2D Rayleigh waves, which is significant for practical applications, such as, the potential application for the grain size measurement with more sensitivity. Meanwhile, it is hoped that these findings will be useful for future studies of 3D Rayleigh wave scattering and that they may lay the groundwork for developing an approach to achieve efficient 2D models which are usefully representative of 3D phenomena.

\subsection{\label{sec:4.2}Prediction of grain noise measured with plane Rayleigh wave excitation }

\begin{figure}[htp]
\centering
\includegraphics[width=1\textwidth]{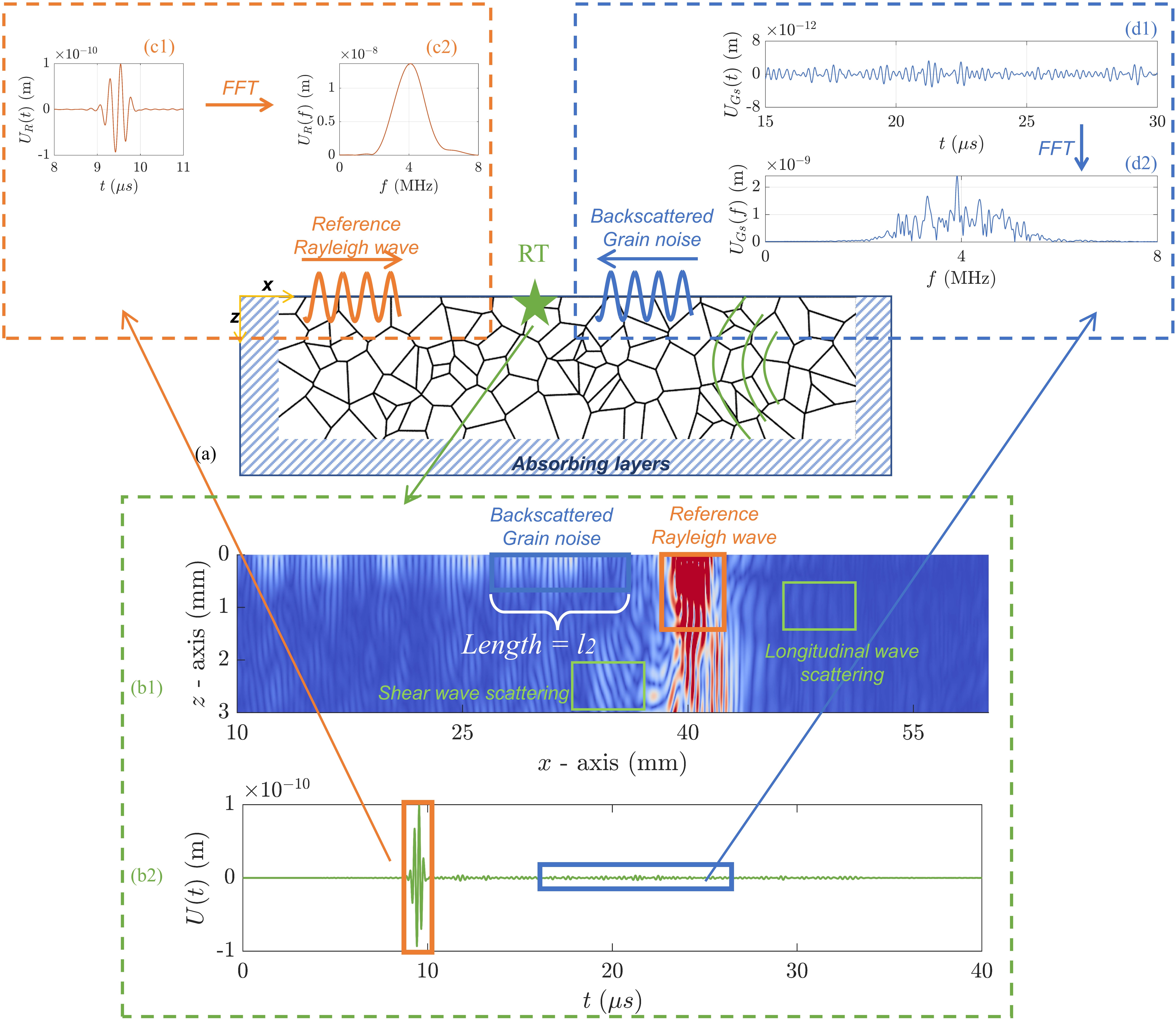}
\caption{\label{fig:7} Example FE modelling of Rayleigh waves propagating on the surface of a polycrystalline material (a) the FE model setup and the simulated fields at (b1) the point `RT' at the time of $t = 27 \mu s$  for Rayleigh wave propagating on the surface of a polycrystal material. (c) the $z-$ displacement for the reference Rayleigh waves (orange rectangle) in the time/frequency domain. (d) the $z-$ displacement for the selected grain within a blue rectangular time window in the time/frequency domain.}
\end{figure}

In this section, the grain noise generated with plane Rayleigh wave excitation is investigated with the numerical method, which is used to compare with the theoretical model given in Sec. \ref{sec:2.3} (Eq. \ref{eq:031}). The models shown in Table \ref{tab:1} are still used in this section. All setup of simulations is similar to that we mentioned above. What is different from the above simulation is that plane Rayleigh waves propagate on the polycrystalline material's surface directly, as shown in Fig. \ref{fig:7}(a). For each case discussed in this section, five realisations with random crystallographic orientations of the model are run.  An incoherent average (rms) is taken over the signals received by the receiver (point `RT' in Fig. \ref{fig:7}(a)). Then, the rms of the  averaged signals will be used for discussion. 

The processing results from the backscattered signals are also indicated in Fig. \ref{fig:7}. Figure \ref{fig:7}(b1) is the Rayleigh wave field in the model after exciting the source nodes with a signal of 1 MHz centre frequency. It can be seen that except for the Rayleigh wave scattering, there are still some bulk wave scattering behaviours. The fact is that the scattered Rayleigh wave is about 100 times stronger than the scattered longitudinal wave, which means the contribution of longitudinal waves is negligible. Meanwhile, we can control the selected time range to reduce the effect of scattered waves caused by shear waves on the final results, which will be discussed next. The signal related to the reference Rayleigh waves and backscattered grain noise caused by the multiple grains is illustrated in Fig. \ref{fig:7}(b2) in the time domain. The respective time/frequency-domain amplitude spectra for the reference signal and backscattered grain noise are displayed in Figs. \ref{fig:7}(c) and (d). The reference signal $U_{\mathrm{ref}}\left( t \right)$ and the backscattered grain noise $U_\mathrm{Gs}\left( t \right)$ are Fourier transformed into the frequency domain to obtain the spectra $U_{\mathrm{ref}}\left( f \right)$ and $U_\mathrm{Gs}\left( f \right)$. The  frequency-dependent normalised grain noise with $j$ realisations is then calculated by 
\begin{equation}
    N^{j}\left( f \right) = U_\mathrm{Gs}^{j}\left( f \right)\big /U_{\mathrm{ref}}^{j}\left( f \right) \quad \text{and} \quad N_{\mathrm{rms}}^{\mathrm{FE}}\left( f \right) = \textbf{RMS} \left[N^{j}\left( f \right)\right]\big /\sqrt{l_2} \textrm{,} 
\end{equation}
where $l_2$ is the length corresponding to the selected grain noise. $N_{\mathrm{rms}}^{\mathrm{FE}}\left( f \right)$ will be used to evaluate the theoretical model result, $N_{\mathrm{rms}}\left( \omega \right)$, in the next. 

\begin{figure}[htp]
\centering
\includegraphics[width=0.6\textwidth]{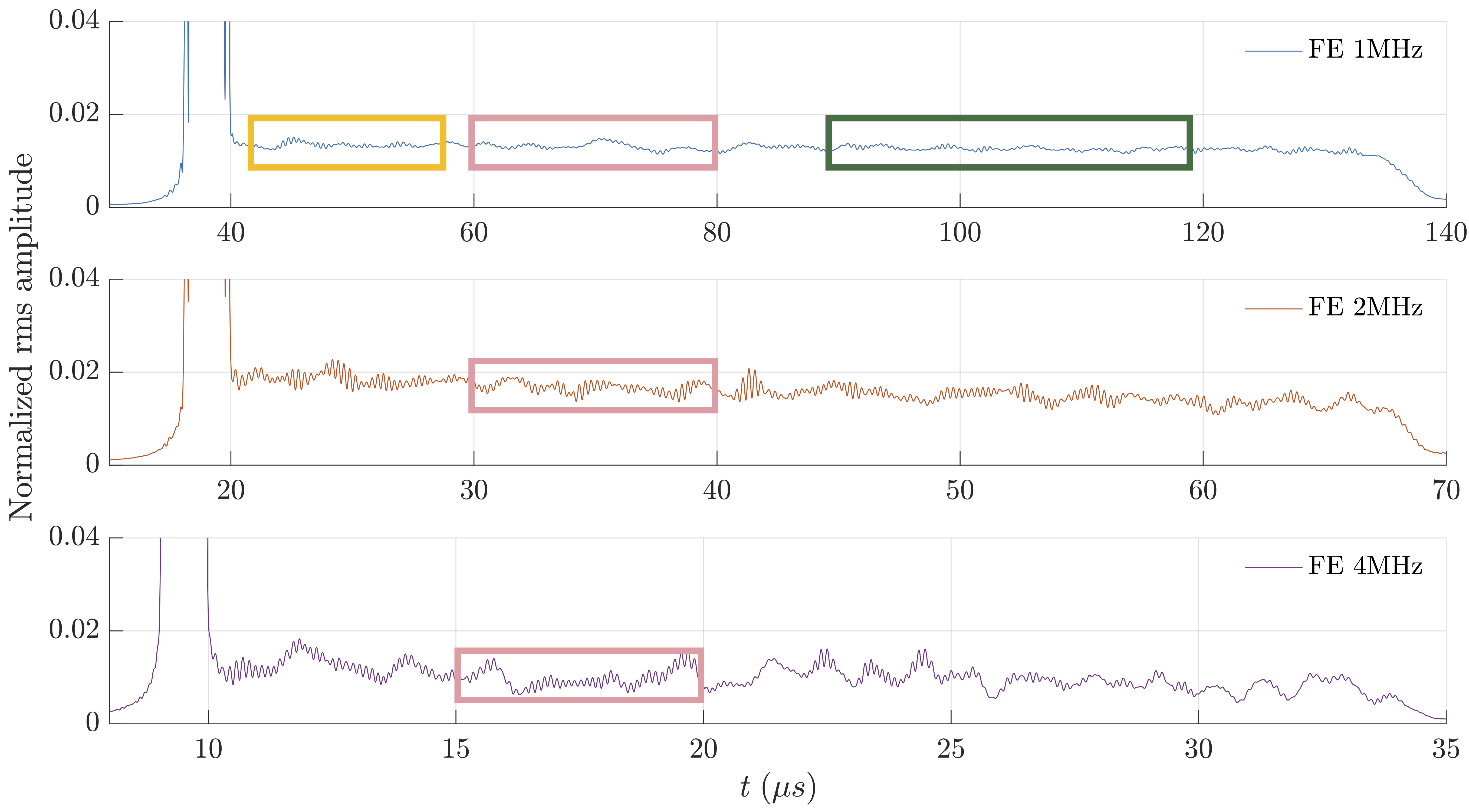}
\caption{\label{fig:8} Spatial grain noise received by the receiver for three different frequencies. The main wave packet corresponding to the reference transmitted wave is clipped where needed to highlight the grain noise.}
\end{figure}

Figure \ref{fig:8} shows the RMS signals in the time domain under 1, 2, and 4 MHz excitation. Aluminium is considered here.  To highlight the grain noise, the reference transmitted waves are clipped in the figure. It can be seen that the grain noise is independent of time at a lower frequency (1MHz), while the grain noise decreases with time at a relatively higher frequency (4MHz), which indicates inherently multiple scattering effects. Therefore, in order to reduce the bulk wave scattering and multiple scattering effects, the FE results received in the appropriate early time range are used later for comparison with the theoretical predictions. For example, in the 1 MHz simulation case, in order to avoid the shear scattering in the very early time range (yellow rectangle) and multiple scattering in the later time range (green rectangle), the appropriate early time range from 60 to 80 $\mu s$ (pink rectangle) will be used to get the grain noise signal. Similarly,  30 $\sim$ 40 $\mu s$ and 15 $\sim$ 20 $\mu s$ are applied  for 2 MHz and 4 MHz, respectively.

\begin{figure}[htp]
\centering
\includegraphics[width=0.9\textwidth]{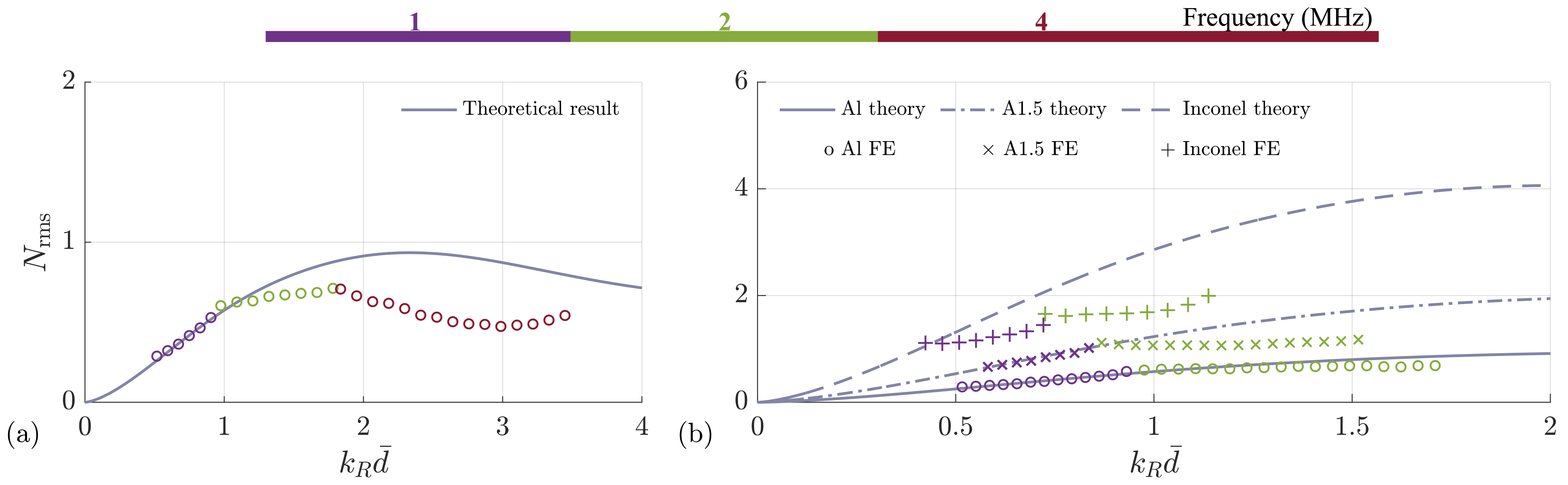}
\caption{\label{fig:9} Comparison of FE and theoretical grain noise, (a) for aluminium material with a larger range of $k_R \Bar{d}$; (b) for different anisotropy indices with aluminium (1.24), A = 1.5 (1.52), and Inconel (2.83), respectively. Note that the y-axis range in (b) is three times of that in (a).}
\end{figure}

Figure \ref{fig:9}(a) illustrates the comparison of the FE and theoretically predicted grain noise with the aluminium material. The centre frequency used in FE simulations is in the range from 1 to 4 MHz. The grey curve shows the theoretical predictions for the aluminium material. The coloured points are the numerical grain noise. From the figure, a good agreement between the two predictions can be seen with a smaller $k_R \Bar{d}$. With the increase of the $k_R \Bar{d}$, a larger discrepancy between the theoretical prediction and numerical results is showing up. It can be explained that the Born approximation is gradually failing and the multiple scattering cannot be neglected, as we discussed in Fig. \ref{fig:8}, with a larger $k_R \Bar{d}$. Meanwhile, we want to emphasise that for only considering the R-R single scattering case with aluminium material (as shown in Fig. \ref{fig:5}(a)), the theoretical model still works well even at $k_R \Bar{d}$ = $2\pi$ (i.e. $d = \lambda_R$, where $\lambda_R$ is the wavelength of Rayleigh waves), while the case with multiple scattering is lost accuracy at $k_R \Bar{d}$ = $1.2$. It means that multiple scattering is an important contribution to the scattering behaviour of Rayleigh waves and is ignored by the theoretical model. 

Furthermore, to observe how the agreement changes with anisotropy, the comparisons for the aluminium, A=1.5, and Inconel materials, which have anisotropy indices of 1.24, 1.52, and 2.83 are performed, as shown in Fig. \ref{fig:9}(b). we note that the y-axis range in (b) is three times of that in (a).  It is clearly demonstrated that the agreement between the theoretical and FE results decreases as the anisotropy index increases. Such results are reasonable because of the Born approximation and multiple scattering, which have been discussed above. 

In addition, the theoretical results are always overestimating the grain noise in larger $k_R \Bar{d}$ regions. As we mentioned before, the Born approximation always has an underestimation for the backscattering amplitude. Therefore, it is not caused by the Born approximation. In fact, when the Rayleigh waves propagate along the surface, the energy of the coherent wave decreases with the increase of the distance far from the observation point in the FE model due to the effect of the multiple scattering. Moreover, it implies the backscattered wave energy from these grains which are far from the observation point is decreasing. The result is that the backscattered energy in different cross sections in the propagation direction will not be equal, which is not like that assumed in the IS approximation. Meanwhile, it should be emphasised that multiple scattering has a larger effect on the accuracy of the theoretical model compared to the Born approximation.

In this section, the rms for single backscattering amplitudes are scattered from single grains with different shapes and random orientations are discussed in Sec. \ref{sec:4.1} and the backward grain noise excited and captured has been predicted with the theoretical model and FE approaches in Sec. \ref{sec:4.2}. Results imply a good agreement between the theoretical prediction and the FE result. Born approximation always has an underestimation of the theoretical backscattering amplitude. In addition, the multiple scattering has little influence on the grain noise level as the anisotropy factor is relatively low and backscattering is weak. However, with a relatively high anisotropy level or a larger normalised frequency, discrepancies are observed indicating the occurrence of multiple scattering. 
Meanwhile, the existence of multiple scattering makes the FE results smaller than that of theory. We note that the ignoring of multiple scattering will result in a larger difference between the theoretical predictions and FE results, compared with the difference caused by the Born approximation. However, the effects of multiple scattering and Born approximation are intertwined, which makes it difficult to quantify how small should  $k_R \Bar{d}$ be for the theoretical model to be valid. Therefore, only a qualitative discussion related to how these two parameters affect the theoretical model has been made.

\section{\label{sec:5}Conclusion}
In this work, we developed a 2D theoretical model for Rayleigh-to-Rayleigh backscattering in untextured polycrystalline materials with equiaxed grains. The model is formulated in the frequency domain based on the Born approximation. A FE model is established to provide relatively accurate reference data for evaluating the approximations of the theoretical model. The comparison of the theoretical and FE results led to various conclusions, mainly including:

1. Good agreement between the FE and theoretical backscattering amplitudes predictions can be seen in the case with only R-R single scattering. With the increase of the anisotropy index, the discrepancy is larger as a result of the use of the Born approximation.

2. The backscattering amplitude is proportional to one and a half power of frequency when the wavelength is comparable to the size of the average grain or much larger than the average grain size, i.e. $k_R \Bar{d} < 1$. The backscattering amplitude is independent of frequency in the case that the wavelength is smaller than the average grain size ($k_R \Bar{d} > 10^{1}$). The quantitative relationship between the normalised frequency and backscattering amplitude is similar to that of 2D bulk wave backscattering behaviour.

3. With the consideration of the weak multiple scattering, the agreement between the theoretical model and FE results is still excellent. The discrepancy seen in the highly scattering case (larger anisotropy index or the wavelength smaller than the average grain size) represents the larger effect of ignoring multiple scattering, rather than the Born approximation, on the backscattering predictions for the theory. 

4. FE is attractive in describing the backscattering noise behaviour as it does not involve some of the important simplifying assumptions included in the theoretical model and therefore can capture multiple scattering which is ignored in the theory.

Generally speaking, we have demonstrated the applicability of our theoretical model to evaluate the backscattering behaviour of Rayleigh waves on a polycrystalline material with single-phase, untextured, and equiaxed grains. We have employed FE simulations as perfectly controlled experiments, where the material properties and configurations are user-defined and accurate, to successfully validate the theoretical predictions. Future studies will be focused on experimental verification, with the aim of utilising this mathematical model for material characterisation in practice. The potential application areas include developing a Rayleigh wave scattering model for polycrystals with elongated grains, performing the experimental inversion of grain size, evaluating the scattering attenuation of Rayleigh waves and characterising the grain size variation in the depth direction.

\printcredits

\section*{Acknowledgements}
    This work was supported by the China Scholarship Council, National Natural Science Foundation of China (Grant No. 92060111). BL gratefully acknowledges the Imperial College Research Fellowship. BL and MH thank the generous funding from the NDE group at Imperial and the EPSRC grant EP/W014769/1.

\appendix
\section{The transformation of the integral region with variable change} \label{sec:appendix}

\begin{figure}[ht]
\centering
\includegraphics[width=0.6\textwidth]{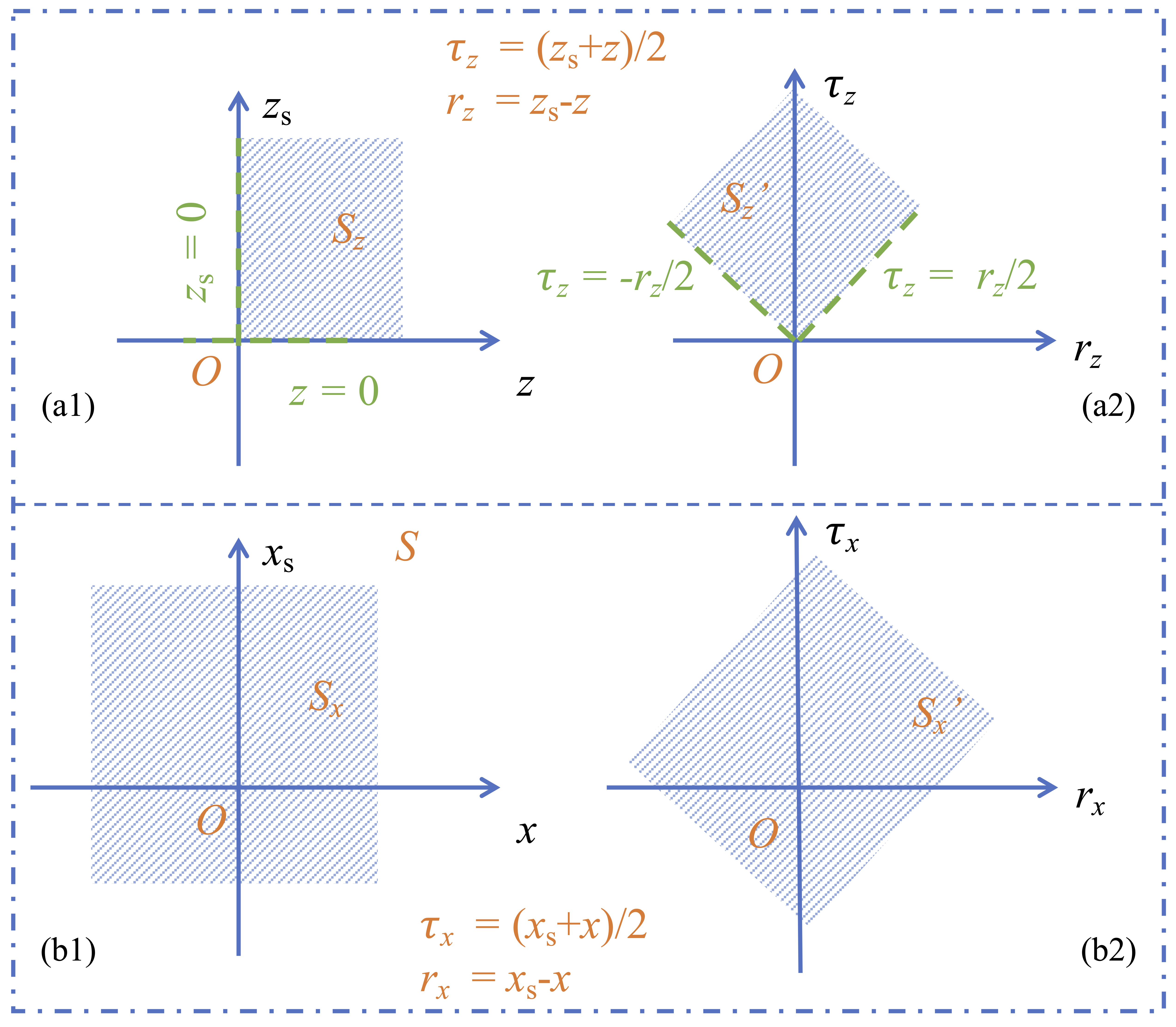}
\caption{\label{fig:A1} Transformation of the integral region in the process of the substitution of the double integral substitution method, (a) the variable change for $z$ and $z_s$ (b) the variable change for $x$ and $x_s$.}
\end{figure}

The change of variables is written as, 
\begin{equation}
   \mathbf{\boldsymbol{\tau}}   = \left(\mathbf{x}_{\mathrm{s}}+\mathbf{x}\right)\big / 2, \quad \mathbf{r} = \mathbf{x}_\mathrm{s}-\mathbf{x}.
\end{equation}
Then the following equation is straightforward,
\begin{equation}
\begin{split}
   &\tau_z   = \left(z_s+z\right) \big / 2, \ r_z = z_s-z,  \\
   &\tau_x = \left(x_s+x\right) \big /2, \ r_x = x_s-x. 
\end{split}
\end{equation}
The limit before the variable change is expressed by,
\begin{equation}
\begin{split}
      &0<z<+\infty,  \ 0<z_s<+\infty, \  -\infty<x<+\infty, \ -\infty<x_s<+\infty
   \end{split}
\end{equation}
The detailed calculation for the limit after the transformation of the integral area is shown in Fig. \ref{fig:A1}. It is clear that the limit change can be written as
\begin{equation}
\begin{split}
    \\ & 0<\tau_z<+\infty, \  -\infty<\tau_x<+\infty, \ -\infty<r_z<+\infty, \ -\infty<r_x<+\infty  
\end{split}
\end{equation}

\pagebreak
\bibliographystyle{elsarticle-num-names}


\end{document}